\documentclass[aps,nofootinbib,amsmath,prd,twocolumn,showpacs,superscriptaddress,groupedaddress]{revtex4-1}

\usepackage[american]{babel}
\usepackage{amsfonts}
\usepackage{amsmath}
\usepackage{amssymb}
\usepackage{epsfig}
\usepackage{latexsym}
\usepackage{paralist}
\usepackage{fancyhdr}
\usepackage{graphicx}
\usepackage[vcentermath]{youngtab}
\usepackage{young}
\usepackage{ytableau}
\usepackage{etex}
\usepackage{braket}
\usepackage{float}
\usepackage{slashed}
\usepackage{xcolor}
\usepackage{soul}
\usepackage{dcolumn} 
\usepackage{physics}
\usepackage{bm}
\usepackage{nicefrac}

\newcommand{\Diff}{{\rm \emph{Diff}}}
\newcommand{\Diffstar}{{\rm \emph{Diff}}^*}
\newcommand{\Weyl}{{\rm \emph{Weyl}}}

\def\bea{\begin{eqnarray}} 
\def\eea{\end{eqnarray}}
\def\be{\begin{equation}} 
\def\ee{\end{equation}} 
\def\ba{\begin{array}}
\def\ea{\end{array}}

\def\be{\begin{equation}}
\def\ee{\end{equation}}
\def\bea{\begin{eqnarray}}
\def\eea{\end{eqnarray}}

\def\Tr{\mbox{Tr}}

\usepackage{amsmath}

\begin{document}

\title{Gravity in $\bm{d=2+\epsilon}$ dimensions and realizations of the diffeomorphisms group}

\author{Riccardo Martini}
\email{riccardo.martini@oist.jp}
\affiliation{
Okinawa Institute of Science and Technology Graduate University, 1919-1, Tancha, Onna,
Kunigami District, Okinawa 904-0495, Japan}

\author{Alessandro Ugolotti}
\email{alessandro.ugolotti@uni-jena.de}
\affiliation{
Theoretisch-Physikalisches Institut, Friedrich-Schiller-Universit\"{a}t Jena,
Max-Wien-Platz 1, 07743 Jena, Germany}

\author{Francesco Del Porro}
\email{fdelporr@sissa.it}
\affiliation{
SISSA, Via Bonomea, 265, 34136 Trieste, Italy}

\author{Omar Zanusso}
\email{omar.zanusso@unipi.it}
\affiliation{
Universit\`a di Pisa and INFN - Sezione di Pisa, Largo Bruno Pontecorvo 3, 56127 Pisa, Italy}

\begin{abstract}
%
We discuss two separate realizations of the diffeomorphism group for metric gravity,
which give rise to theories that are classically equivalent, but quantum mechanically distinct.
We renormalize them in $d=2+\epsilon$ dimensions, developing
a new procedure for dimensional continuation of metric theories and highlighting connections
with the constructions that previously appeared in the literature.
Our hope is to frame candidates for ultraviolet completions of quantum gravity in $d>2$ and give some perturbative
mean to assess their existence in $d=4$, but also to speculate on some potential obstructions in the continuation
of such candidates to finite values of $\epsilon$. 
Our results suggest the presence of a conformal window in $d$ which seems to extend to values higher than four.
\end{abstract}

\maketitle

\renewcommand{\thefootnote}{\arabic{footnote}}
\setcounter{footnote}{0}

\section{Why gravity in $\bm{d=2+\epsilon}$}\label{sect:introduction}

The regularization of a quantum theory of Einstein-Hilbert gravity leads to nonrenormalizable divergences
beyond the first order in the loop expansion \cite{Goroff:1985th,vandeVen:1991gw}.
The problem arises because the perturbative coupling, Newton's constant,
is dimensionful and consequently leads to a non-power-counting renormalizable expansion in $\hbar$.
For this reason, it is generally understood that one should re-arrange the perturbative expansion
in powers of some ultraviolet (UV) energy scale, which could be the Planck mass itself, and that
quantum Einstein-Hilbert gravity can be seen at most as an effective theory below such scale \cite{Donoghue:1994dn}.

A different perspective on the problem is achieved by noticing
that Newton's constant, $G$, is not dimensionful for every spacetime dimension $d$,
hinting that the theory could be perturbatively renormalizable for some value of $d$.
In fact, not only $G$ becomes dimensionless in $d=2$ spacetime dimensions,
but it also has an asymptotically free beta function, $\beta_G \propto - G^2$,
meaning that, in principle, one could obtain consistent predictions
from perturbation theory, that are valid up to arbitrarily high energies \cite{Kawai:1989yh}.
We refer to $d=2$ as the \emph{critical dimension} of the Einstein-Hilbert action.

This notion is not particularly useful for the physically interesting $d=4$ case,
unless one realizes that in $d=2+\epsilon$ dimensions one can trade the perturbative expansion in $G$
for an expansion in the parameter $\epsilon$.
Re-instating the canonical mass dimension of Newton's constant, one finds that its
renormalization group (RG) running, in units of an RG scale $\mu$, is $-\epsilon G+ \beta_G \propto - G^2$;
as a consequence there is a scale invariant value $G^*\sim {\cal O}(\epsilon)$,
arising as a fixed point solution of $\beta_G=0$.
The limit $\epsilon\to 0$ has been used to reproduce some results of $2d$ quantum gravity \cite{Kawai:1992np,Codello:2014wfa},
that were previously obtained by other means \cite{Polyakov:1987zb}, but
we anticipate from the main discussion of this paper that the precise value of the proportionality constant in $\beta_G$
has a prominent role in further clarifying this limit\footnote{It is worth mentioning that other approaches try to solve the renormalizability of a quantum theory of metric degrees of freedom by modifying the Einstein-Hilbert action. We refer to \cite{Steinwachs:2020jkj} for an overview of these alternatives and their consequences.}.

The found fixed point $G^*$ has some interesting and useful properties,
because of its UV nature \cite{Kawai:1993mb}.
A simple analysis reveals that it separates the regions
$G<G^*$ and $G>G^*$, which are driven by the RG to the infrared (IR) Gau{\ss}ian
and strongly-interacting limits, respectively.
The UV fixed point in $d=2+\epsilon$ is almost as good as asymptotic freedom:
in the UV the theory always remains consistent by ``approaching'' a nonperturbative
interacting limit instead of becoming asymptotically free.
In the IR, instead, the theory could flow to the Gau{\ss}ian phase and be described
by the aforementioned effective theory, which would see its UV completion as ``hidden''
by the UV energy scale at which Gau{\ss}ian perturbation theory breaks.
A theory with a UV fixed point and a finite number of relevant RG directions is known
as asymptotically safe \cite{Weinberg:1980gg}.

The idea that gravity could be asymptotically safe in $d\geq 2$ and,
most importantly, in $d=4$ has been conjectured a long time ago already by Weinberg \cite{Weinberg:1976xy}.
The existence of the nontrivial UV fixed point ${\cal O}(\epsilon)$
\emph{guarantees} that, at least for infinitesimally small $\epsilon$, there is a UV completion.
This has given a compelling reason to push forward the investigation of the asymptotic safety conjecture,
which has received increasing attention over the past few decades \cite{Bonanno:2020bil}.

Hystorically, after some first papers that pushed forward the status of the conjecture through
the use of perturbation theory and the $\epsilon$-expansion, e.g.~\cite{Jack:1990ey,Aida:1996zn},
most of the literature has eventually settled on the use of a nonperturbative method, known as functional RG,
to test the validity of the conjecture as pioneered in Refs.~\cite{Reuter:1996cp} and \cite{Souma:1999at}
by Reuter and Souma, respectively.
However, the nonperturbative approach comes at a price: the Wetterich equation,
which governs the functional RG flow \cite{Wetterich:1992yh},
comes with a more severe scheme dependence, if compared with massless renormalization schemes
such as dimensional regularization and minimal subtraction
(see appendix \ref{sect:quadratic} for some further comments
on scheme dependence in dimensional regularization and the role of quadratic divergences).
This scheme dependence has the disadvantage of mixing, a bit disturbingly, with both gauge-
and parametrization-dependence of the path-integral \cite{Souma:2000vs}, making unclear which are the physical
predictions of the theory in terms of observables \cite{Donoghue:2019clr, Falkenberg:1996bq}, even though
the problem can somehow be relaxed by going on-shell \cite{Benedetti:2011ct,Benedetti:2015zsw}.
The potential danger, which is faced by the practitioners of the asymptotic safety conjecture,
is that not having clear gauge-independent results undermines the cumulative work towards the
proof of the conjecture itself, and delays further results on other pressing issues
such as, for example, unitarity \cite{Nink:2015lmq}.

Still, we, the authors, are far from having a negative stance towards asymptotic safety,
even after mentioning some of the weaknesses of the modern approach.
Our humble opinion on the conjecture is that the pursuit based on the functional RG
should be coupled with another one that gives less scheme-dependent results, even at the price of nonperturbativity,
and these two searches should come together convincingly.
For this reason, we want to apply the perturbative framework, that we briefly outlined at the beginning,
and resurrect the aged discussion on gravity in $d=2+\epsilon$ dimensions.
Therefore, we follow Weinberg's original idea \cite{Weinberg:1980gg},
and assume that the fixed point of a putative asymptotically free theory
of gravity comes from the continuation to $d=4$ of the asymptotically free theory in $d=2$.
The obvious limitation is that this requires an extrapolation to $\epsilon=2$, which is naive at best and dangerous
at worst, but, fortunately, the results based on perturbation theory should be weighted with the body of work coming
from nonperturbative methods, as we are going to see in the following.
It is important to mention that our idea of using information from perturbation theory is far from new \cite{Falls:2017cze},
although we seem to occasionally come to slightly different conclusions.

One immediate advantage of resorting to perturbation theory is that, even if we start from the same ``classical'' bare theory,
we can identify two \emph{distinct} realizations of the gravitational path-integral based on two isomorphic, but inequivalent,
realizations of the diffeomorphism group. We refer to the two models resulting from these parametrizations
as Einstein's and Unimodular-Dilaton gravity, respectively, following a nomenclature introduced in Ref.~\cite{Gielen:2018pvk}.
The two realizations, which we discuss in the next section,
have often been interpreted as related to the parametrization dependence of the results \cite{Demmel:2015zfa}.
Instead, we explicitly show that each realization does not depend on gauge and other parameters on-shell.
Several ideas have surfaced when revisiting older results with modern eyes,
and we clarify them along the way.

The rest of the paper is structured as follows:
in Sect.~\ref{sect:diff} we discuss the two isomorphic realizations of the diffeomorphisms group and the relative actions;
in Sect.~\ref{sect:c} we explain a three steps procedure
on how to dimensionally continue the theories for the applying regularization methods;
in Sects.~\ref{sect:jj}~and~\ref{sect:ak} we show the leading renormalization of the two different actions;
in Sect.~\ref{sect:d2} we expand our discussion on the role of the conformal mode and the two dimensional limit;
in Sect.~\ref{sect:conclusion} we attempt a conclusion and speculate on the results;
in appendix~\ref{sect:quadratic} we briefly touch the topic of quadratic divergences in dimensional regularization.

\section{${\Diff}$ vs ${\Diffstar}$}\label{sect:diff}

As starting point of this discussion we use the familiar (Euclidean) Einstein-Hilbert action for \emph{Einstein gravity} (EG)
in $d$-dimensions
\begin{equation}\label{eq:einstein-hilbert}
\begin{split}
 S_E[g] &= \int{\rm d}^dx \sqrt{g}\Bigl\{ g_0-g_1 R\Bigr\}\,,
\end{split}
\end{equation}
in which we introduced the curvature scalar $R$ and two couplings, $g_0$ and $g_1$, containing the traditional cosmological
and Newton's constants. The strength of the gravitational interaction is weighted by $G$,
which we choose to be $g_1=G^{-1}$, avoiding any unnecessary normalization.
Due to the covariant nature of all the elements appearing in \eqref{eq:einstein-hilbert}, the action is
manifestly invariant under diffeomorphisms. Infinitesimal diffeomorphisms act as changes of coordinates
and are generated by a vector field,
$x^\mu \to x^\mu +\xi^\mu(x)$, resulting in a transformation of the metric
\begin{equation}\label{eq:diff-action}
\begin{split}
 \delta_\xi g_{\mu\nu} &= {\cal L}_\xi g_{\mu\nu} = \nabla_\mu \xi_\nu+ \nabla_\nu \xi_\mu\,.
\end{split}
\end{equation}
Since the composition of two transformations is still a transformation, the diffeomorphisms form a group which we denote ${\Diff}$.
The algebra is obviously closed
\begin{equation}\label{eq:diff-algebra}
\begin{split}
 \left[\delta_{\xi_1},\delta_{\xi_2}\right] &= \delta_{\left[\xi_1,\xi_2\right]}\,,
\end{split}
\end{equation}
where on the right hand side the commutator denotes the standard Lie brackets of two vector fields.

Following Ref.~\cite{Gielen:2018pvk}, we now define a slightly more general action
which goes under the name of
\emph{dilaton gravity} (DG) action. We begin by parametrizing the metric
as a dilatonic factor times another metric, $g_{\mu\nu}=\Omega^2 \tilde{g}_{\mu\nu}$. This results in the dilaton
action, $S_D[\Omega,\tilde{g}] \equiv S_E[g]$, which has an additional local Weyl symmetry
caused by the possibility of rescaling the two factors $\Omega^2$ and $\tilde{g}_{\mu\nu}$
while leaving $g_{\mu\nu}$ invariant. We further parametrize $\Omega= \varphi^{\nicefrac{2}{d-2}}$
to get
\begin{equation}\label{eq:dilaton}
\begin{split}
 S_D[\varphi,\tilde{g}] =& -g_1\int {\rm d}^d x \sqrt{\tilde{g}}\Bigl\{\varphi^2 \tilde{R}
 + 4\frac{d-1}{d-2} \tilde{g}^{\mu\nu}\partial_\mu\varphi\partial_\nu\varphi\Bigr\}
 \\&
 + g_0 \int {\rm d}^d x \sqrt{\tilde{g}}\Bigl\{ \varphi^{\nicefrac{2d}{d-2}} \Bigr\} 
\,,
\end{split}
\end{equation}
with $\tilde{R}=R(\tilde{g},\partial \tilde{g})$.
An additional rescaling of the field $\varphi \to \alpha \varphi$ by the constant $\alpha=\nicefrac{(d-2)}{8(d-1)}$
can be used to normalize the kinetic term of $\varphi$; 
it brings the first line to a familiar conformal invariant action with nonminimal coupling in $d$ dimensions
(coupled with the so-called Yamabe operator) \cite{Chernicoff:2018apt}.
For $d>2$ the required rescaling is purely imaginary and is generally associated to the
\emph{conformal mode instability} because of the wrong sign of the kinetic term of \eqref{eq:dilaton},
while for $1<d<2$ there is no such problem. We defer the discussion on the case $d=2$ to Sect.~\ref{sect:d2}.

By construction, the dilaton action \eqref{eq:dilaton} has two symmetries: a diffeomorphism invariance in which the two fields
transform independently
\begin{equation}\label{eq:diff-dilaton-action}
\begin{split}
 &\delta_\xi \tilde{g}_{\mu\nu} = {\cal L}_\xi \tilde{g}_{\mu\nu} = \tilde{\nabla}_\mu \xi_\nu+ \tilde{\nabla}_\nu \xi_\mu\,,
 \\
 &\delta_\xi \varphi = {\cal L}_\xi \varphi = \xi^\mu \partial_\mu \varphi\,,
\end{split}
\end{equation}
(indices are raised and lowered through $\tilde{g}_{\mu\nu}$ and its inverse),
and a Weyl invariance on local rescalings with respect to an arbitrary scalar function
that leave the original $g_{\mu\nu}$ invariant.
Infinitesimally, the Weyl transformation acts as
\begin{equation}\label{eq:weyl-dilaton-action}
\begin{split}
 & \delta^{w}_\omega  \tilde{g}_{\mu\nu}= 2 \omega \tilde{g}_{\mu\nu}\,,
 \\
 & \delta^{w}_\omega \varphi = \left(1-\nicefrac{d}{2}\right) \omega \varphi\,,
\end{split}
\end{equation}
for some infinitesimally small scalar function $\omega$.
The complete symmetry group of \eqref{eq:dilaton} is a semidirect product of the two subgroups
\begin{equation}\label{eq:full-dilaton-symmetry}
\begin{split}
 \Diff \ltimes {\Weyl}
\end{split}
\end{equation}
because of the nontrivial action of ${\Diff}$ on $\Weyl$.

It is clear that the symmetry group of the dilaton action \eqref{eq:dilaton} is enhanced because
of the redundancy introduced by the combination $g_{\mu\nu}=\Omega^2 \tilde{g}_{\mu\nu}$.
The obvious way to get rid of both the redundant symmetry and the scalar degree of freedom $\Omega$
is to take $\Omega=1$, which breaks down \eqref{eq:full-dilaton-symmetry} to the original ${\Diff}$ group
and \eqref{eq:dilaton} to \eqref{eq:einstein-hilbert}.
However, there is an alternative way to break the symmetry group, which involves assuming that the metric theory
based on $\tilde{g}_{\mu\nu}$ in \eqref{eq:dilaton} is \emph{unimodular}. In the unimodular theory we require
any possible metric $\tilde{g}_{\mu\nu}$ to obey the condition $\sqrt{\tilde{g}}=v$, where
$v$ is some fixed volume $d$-form (oftentimes just chosen $v=1$).

We refer to the unimodular version of \eqref{eq:dilaton} as \emph{unimodular-dilaton gravity} (UDG),
and its action is simply \eqref{eq:dilaton}, with the restriction that the metric $\tilde{g}_{\mu\nu}$ is unimodular.
The symmetry group of the unimodular version of \eqref{eq:dilaton} is the subgroup of \eqref{eq:full-dilaton-symmetry}
that contains all the transformations that leave a given $v$ invariant.
Infinitesimally, it is generated by the combined transformations $\delta^*=\delta_\xi+\delta_\omega^w$,
from \eqref{eq:diff-dilaton-action} and \eqref{eq:weyl-dilaton-action},
for which $\delta^* \sqrt{\tilde{g}}=0$,
thus preventing $v$ from any transformation.
Of course, the generators $\xi^\mu$ and $\omega$ are not independent:
in fact, one way to think at the transformations $\delta^*$ is as standard diffeomorphisms $\delta_\xi$,
which are then compensated by a Weyl transformation with $\omega = \nicefrac{(d-2)}{2d}\,\tilde{\nabla}\cdot \xi$
restoring the original volume element.
Therefore, $\delta^*$ only depends on the vector $\xi^\mu$, hence we adopt the notation $\delta^* \to \delta^*_\xi$.
The action of $\delta^*_\xi$ is thus
\begin{equation}\label{eq:diff-unimodular-dilaton-action}
\begin{split}
 &\delta^*_\xi \tilde{g}_{\mu\nu} = {\cal L}_\xi \tilde{g}_{\mu\nu} -\frac{2}{d}\tilde{g}_{\mu\nu} \tilde{\nabla}\cdot \xi
 \,,
 \\
 &\delta^*_\xi \varphi = {\cal L}_\xi \varphi +\frac{d-2}{2d} \varphi\, \tilde{\nabla}\cdot \xi
 \,.
\end{split}
\end{equation}
Obviously, the unimodular transformations are a subgroup of
\eqref{eq:full-dilaton-symmetry}, which can be though of as a
symmetry-breaking pattern coming from the quotient of \eqref{eq:full-dilaton-symmetry}
with the equivalence relation $\delta^* \sqrt{\tilde{g}}=0$. We denote it as ${\Diffstar}$:
\begin{equation}\label{eq:breaking}
\begin{split}
 \Diff \ltimes {\Weyl}  ~
\overset{\delta^*\sqrt{\tilde{g}}=0}{\longrightarrow} ~
 {\Diffstar}\,.
\end{split}
\end{equation}
Importantly, the new group is isomorphic to ${\Diff}$ itself,
${\Diffstar}\simeq{\Diff}$, using the relation
\begin{equation}\label{eq:diffstar-algebra}
\begin{split}
 \left[\delta^*_{\xi_1},\delta^*_{\xi_2}\right] &= \delta^*_{\left[\xi_1,\xi_2\right]}\,,
\end{split}
\end{equation}
which can be proven with a bit of work.
This could be foreshadowed by the one-to-one correspondence between transformations and vector fields $\xi^\mu$,
together with the fact that, modulo normalizations, two vectors combine into a third one only through Lie brackets.\footnote{%
The result is less trivial than it looks, though,
because a different breaking pattern can lead to yet another symmetry group \cite{Gielen:2018pvk}.
One such possibility is that we are left with a semidirect product of unimodular diffeomorphisms and $\Weyl$ transformations.
}
Of course, in this discussion we are identifying the groups with their algebras,
so our statements hold, realistically, only at the infinitesimal level and the groups might differ globally.

Having identified the larger symmetry group in the breaking pattern \eqref{eq:breaking} leading to ${\Diffstar}$,
we find convenient to write a more general action for the UDG theory, in which the condition of $\Weyl$ invariance is
relaxed
\begin{equation}\label{eq:unimodular-dilaton}
\begin{split}
 S_{U}[\varphi,\tilde{g}] =& -g_1\int {\rm d}^d x \sqrt{\tilde{g}} \Bigl\{\varphi^2 \tilde{R}
 + 4\frac{d-1}{d-2} \tilde{g}^{\mu\nu}\,\partial_\mu\varphi\,\partial_\nu\varphi\Bigr\}
 \\&
 +  \int {\rm d}^d x \sqrt{\tilde{g}}\, V (\varphi) + q\int {\rm d}^d x \sqrt{\tilde{g}}\, \varphi\, \tilde{R}
\,,
\end{split}
\end{equation}
in which we introduced a function, $V(\varphi)$, that plays the role of cosmological constant
and a topological charge $q$ that couples $\varphi$ to the curvature scalar in a way familiar to
Liouville field theory \cite{Levy:2018bdc}.

There are important reasons behind the chosen parametrization.
The charge $q$ breaks classical Weyl invariance because it gives a nonzero value to the trace of
the stress energy tensor of $\varphi$ over $\tilde{g}_{\mu\nu}$, that is, $T=T^\mu{}_{\mu}=-2q^2 R$.
The value $q$ can, however, be chosen so that the quantum Weyl anomaly is zero
$\langle T\rangle=-2q^2 R+ T^{\rm rad}=0$, where the second term comes from radiative corrections \cite{David:1988hj}.
This can be done consistently order by order in perturbation theory, as shown in Ref.~\cite{Kawai:1995ju}.
The quantum theory is thus invariant under Weyl transformations only if $q$ is chosen appropriately,
and the function playing the operator $V(\varphi)$, becomes $V(\varphi)=g_0 \varphi^{\nicefrac{2d}{d-2}}$.
The reason why it is important to restore Weyl invariance is that the action \eqref{eq:unimodular-dilaton},
in the words of Ref.~\cite{Aida:1994zc}, is invariant under ``volume preserving diffeomorphisms'',
but at a Weyl-invariant critical point these are promoted to ``full diffeomorphisms.''
In our discussion, we identify the volume preserving transformations as unimodular transformations,
and the full diffeomorphisms group as ${\Diffstar}$, which is correctly isomorphic to $\Diff$, though not
exactly the same, as we previously argued following Ref.~\cite{Gielen:2018pvk}.
Conformal invariance should emerge as a symmetry of the (scale invariant) RG fixed point.
Notice also the difference of this approach with the traditional unimodular gravity realization without the dilaton, in which
the cosmological constant emerges as an integration constant of the equations of motion \cite{Benedetti:2015zsw}.

\begin{figure}
\includegraphics[width=0.4\textwidth]{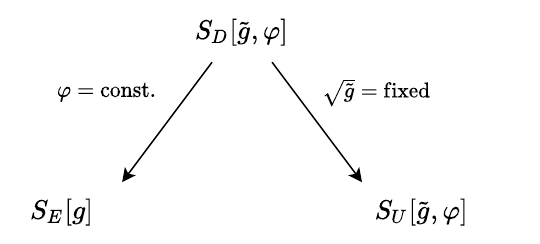}
\caption{Summary of the theories and the breaking patterns relating them. \label{fig:theories}}
\end{figure}

A more general analysis could be implemented by promoting the operators related to the curvature scalar to function too,
as suggested in Ref.~\cite{Martini:2018ska}, which would contain both $g_1$ and the topological charge.
In this case the breaking \label{eq:diffstar-breaking} would be entirely realized at the fixed point starting from a dilatonic theory,
but we leave this possibility open for now.
A simple visual summary of the breaking patterns relating the three actions is given in Fig.~\ref{fig:theories},
which depicts an upper wedge of a more general pattern described in Ref.~\cite{Gielen:2018pvk}.

For the rest of the paper we focus on the Einstein and unimodular-dilaton realizations,
with actions \eqref{eq:einstein-hilbert} and \eqref{eq:unimodular-dilaton}, respectively.
The symmetry groups of the two theories are isomorphic, so the natural question
is whether they actually are the same physical model or not.
From the Hamiltonian analysis emerges that the two models propagate the same degrees of freedom \cite{Gielen:2018pvk},
so it is fair to say that classically the Einstein's and unimodular-dilaton
gravitational models are equivalent.
What is less clear is if the two models are quantum mechanically equivalent.
One way to investigate this difference is to inspect the beta function of Newton's constant in each realization.
From a general scaling analysis, see for example Ref.~\cite{Codello:2014wfa}, we know that the beta function must be of the form
\begin{equation}\label{eq:general-beta}
\begin{split}
 \beta_G = \frac{c}{24\pi} G^2
\,,
\end{split}
\end{equation}
in which we denoted with $c$ the central charge, which is a number of particular importance
in the context of two-dimensional conformal field theory (CFT)
because it counts the effective number of degrees of freedom
that are integrated in the path-integral. In general, $c$ will be the sum of two contributions,
one coming from the matter fields
and the other coming from the gravitational degrees of freedom.

As discussed in Refs.~\cite{Kawai:1989yh,Falls:2017cze}, the value of the central charge for two dimensional quantum gravity is scheme dependent. In particular, coupling the theory with matter allows for the introduction of operators of different mass dimensions and the prescriptions to go on-shell increase since it is possible to solve the equations of motion with respect to different operators. The coefficient of the beta function for Newton's constant is then shifted by a value proportional to the mass dimensions of the operator solved by the equations of motion \cite{Falls:2017cze}.

In absence of matter,though, possibility are much more restricted and the allowed schemes are determined by the realization of the symmetry chosen for the gravitational sector solely.
We have that in $d=2$ Einstein's \eqref{eq:einstein-hilbert} leads to $c=-19$ \cite{Jack:1990ey},
while unimodular-dilaton's \eqref{eq:unimodular-dilaton} might also lead to $c=-25$ \cite{Aida:1996zn}.
Evidently, in both cases $c<0$, implying that a fixed point $g^*\sim{\cal O}(\epsilon)$ exists,
but it is undeniable that the quantum mechanical effects are different.
Notice that, in the past, this difference was often attributed to a parametrization dependence 
(exponential vs linear background split of the metric),
and more often than not it was believed that the second result is the ``correct'' one
because it reproduces results from string theory and Liouville gravity
in the limit $\epsilon\to 0$ \cite{Distler:1988jt,Kawai:1989yh}.
By generalizing results for both theories, in the following sections
we show that the difference has nothing to do with parametrizations,
rather the symmetry plays a much bigger role.
A notable exception of a paper, that partly discusses the two models
as two similar but independent realizations, each with its own central charge
is Ref.~\cite{Nink:2015lmq}.

\section{Three steps continuation in $\bm{d}$}\label{sect:c}

The analytic continuation to $d=4$ of a $d=2+\epsilon$ computation requires the extrapolation
to $\epsilon=2$, which is obviously a big limitation of the perturbative approach
that works well at small $\epsilon$.
Nevertheless, the two dimensional limit, $\epsilon \to 0$,
can be used to give theoretical arguments against or in favor of either
realization of ${\Diff}$ described in the previous section.

The strategy that we adopt for dealing with the regularization of
either of the theories discussed in Sect.~\ref{sect:diff}
is dimensional regularization with modified minimal subtraction ($\overline{\rm MS}$) of the divergences
close to $d=2$, implying that we subtract the poles $\nicefrac{1}{d-2}$
and a small finite part after analytic continuation of the results in the dimensionality.
There are several difficulties that arises with a naive application of $\overline{\rm MS}$ in gravity.
The first and most prominent one is that several tensor contraction
also have $d$-dependent outcomes, such as $g_\mu{}^\mu=d$,
which might change the finite part of the subtractions when multiplying a pole or,
in the worst case, entirely remove a divergence.
This obviously could make ambiguous the status of some divergences.

In order to lift any ambiguity we define three fundamental steps for dimensional continuation and regularization
of the metric theories, while keeping in mind that the desired outcome is extrapolation \emph{above} $d=2$.
The procedure will unfortunately imply a proliferation of $\epsilon$-like symbols.

\begin{itemize}

\item The first step is that we actually continue any Feynman diagram\footnote{%
More precisely, covariant Feynman diagram, since in Sects.~\ref{sect:jj} and \ref{sect:ak}
we use covariant heat kernel methods to compute the effective action.
}
in the dimension as $d=2 \to d=2-\zeta$;
we stress that, for the moment, $\zeta \neq -\epsilon$ which was previously introduced.
This means that, whenever an integration measure appears, we promote it as ${\rm d}^2q \to {\rm d}^{2-\zeta}q$.
For $\zeta>0$ the diagrams that are relevant for perturbation theory converge
(generally for $\Re \zeta >0$),
but, equally importantly, the conformal mode of the metric is stable
(see also the discussion in Sect.~\ref{sect:d2}; in short it means that the constant in front of
the kinetic term of $\varphi$ in \eqref{eq:dilaton} is positive).
We elect the regime $\zeta >0$ as the one in which
we make computations of the radiative corrections; divergences thus appear as poles $\nicefrac{1}{\zeta}$
that must be subtracted with opportune counterterm operators,
and their coefficients assemble into beta functions of renormalized couplings.

\item The second step is that, any time a tensorial contraction returns the dimension of spacetime,
we simply denote such dimension with $d$ and treat it as a parameter.
The simplest example is of course $g_\mu{}^\mu=\delta_\mu^\mu=d$.
In particular, and this is very important, we do \emph{not} substitute $d=2$ \emph{nor} $d=2-\zeta$ or $d=2+\epsilon$
when computing divergences, though, either limit can be taken later if needed.
This has the advantage that $d$ appears parametrically in our computations,
much like $N$ appears when computing the renormalization of an $SU(N)$ gauge theory.
Similarly to gauge theories, by setting $d=2$ or $d=4$ the metric fluctuations have the expected degrees of freedom
in a given dimension.

\item The third step is the one of continuing the results to $d>2$.
This is achieved by continuing $\zeta= -\epsilon$ with $\epsilon>0$, that is the forbidden region $\zeta<0$,
which explains why we keep it separate from the process of dimensionally regularizing the theory.
What we imply in this last step is that only \emph{after} having regulated and renormalized
the model, and obtained a beta function such as \eqref{eq:general-beta},
we then continue  to $d=2+\epsilon>2$ dimensions by noticing that the coupling $G$ must have negative mass dimension.
This step introduces the dimensionless coupling through the replacement
$G \to G \mu^{-\epsilon}$, where $\mu$ is the RG scale, effectively measuring
the coupling constant in units of $\mu$. The net effect of this last step is that
the beta function $\beta_G$ acquires the scaling term $\epsilon G$ as explained in Sect.~\ref{sect:introduction}.

\end{itemize}

In principle, the three above steps leave us with two independent parameters related to the meaning of dimension $d$,
which can or cannot be identified, according to necessity.
A summary of our general strategy is the following: we eliminate poles in $\zeta$ coming from diagrams
entirely though $\overline{\rm MS}$-like subtraction (first step),
so we can express the beta functions as $d$-dependent objects (second step),
that we continue to $d>2$ (third step).
The final result in an $\epsilon$-expansion series that has $d$-dependent coefficients.
With these steps in mind, it should be clear that
one can investigate the two dimensional limit by taking $d=2+\epsilon$ and $\epsilon\to 0$,
but can also estimate the four dimensional limit by taking $d=4$
and extrapolating $\epsilon$ to $\epsilon\to 2$.
This procedure is clarified by the explicit examples in the next two sections.

The advantage of our procedure is that it breaks down the problematic dimensional continuation
of a gravitational theory in manageable steps, which have a range of validity that is under control
and can be discussed separately.
For example, the first step ensures that the conformal mode is actually stable and makes the integrals converge,
thus picking up a specific vacuum \cite{Hawking:1978jn}, which is different from, say,
the one in Refs.~\cite{Morris:2018mhd,Mitchell:2020fjy,Morris:2020blt,Kellett:2020mle}.
The second step mostly ensures that we have the expected number of propagating degrees of freedom in a given dimension.
Our opinion is that the potentially dangerous step is the third one,
because it involves the continuation above $d=2$ by means of the
$\epsilon$-expansion in a regime in which the theory does not converge, strictly speaking,
but we defer a more thoughtful discussion of this point for the speculative part of the conclusions.

\section{Einstein's action}\label{sect:jj}

We begin by considering the Einstein-Hilbert action \eqref{eq:einstein-hilbert}.
The path-integral can be constructed using the background field method by splitting the metric as
\begin{equation}\label{eq:linear-split}
\begin{split}
 g_{\mu\nu} &= \overline{g}_{\mu\nu} + h_{\mu\nu}
+\frac{\lambda}{2} h_{\mu\rho} \overline{g}^{\rho\theta} h_{\theta\nu} +{\cal O}(h^3)
\,,
\end{split}
\end{equation}
in which $\overline{g}_{\mu\nu}$ is an arbitrary background, $h_{\mu\nu}$ are the fluctuations
to integrate over, and $\lambda$ is an arbitrary parameter to test the parametric dependence of our results.
In principle, many more parameters are hidden in the ${\cal O}(h^3)$ terms,
but we choose to highlight only $\lambda$ to explicitly test the parametric dependence.\footnote{%
There are two main reasons for our choice:
on the one hand, $\lambda$ is the simplest nontrivial parameter that actually influences our computation,
and, on the other hand, it contains the case $\lambda=1$, which matches the exponential parametrization
to the quadratic order.
Independence on $\lambda$ thus translates into independence from the parametrization.
For a general parametrization see Ref.~\cite{Gies:2015tca}.
}
The split allows us to preserve manifest covariance under the background version of \eqref{eq:diff-action},
in other words $\delta_\xi \overline{g}_{\mu\nu} = {\cal L}_\xi \overline{g}_{\mu\nu}$
(while $h_{\mu\nu}$ transforms as a standard symmetric tensor),
which is sometimes known as \emph{background symmetry}.

However, the background symmetry is not the one that we have to gauge-fix, since, in fact, we want to preserve it.
The gauge-fixing must fix \eqref{eq:diff-action} seen as a transformation of $h_{\mu\nu}$
at fixed arbitrary $\overline{g}_{\mu\nu}$, which is nonlinear because of the nonlinearity
of the right hand side of \eqref{eq:linear-split}.
We can reconstruct the correct transformation on $h_{\mu\nu}$ order-by-order in $h_{\mu\nu}$ itself by inverting the following relation  (parentheses indicate symmetrization of the indices):
\begin{equation}
\begin{split}
 g_{\nu\rho}\nabla_\mu \xi^{\rho}+g_{\mu\rho}\nabla_\nu \xi^{\rho} &=
 \delta_\xi h_{\mu\nu} +\lambda h_{(\mu\rho} \overline{g}^{\rho\theta} \delta_\xi h_{\theta\nu)}
 +{\cal O}(h^3)
\,.
\end{split}
\end{equation}
Using \eqref{eq:linear-split} on metrics and connections on the left hand side, we find
\begin{equation}\label{eq:hmunu-transf-linear}
\begin{split}
 \delta_\xi h_{\mu\nu} =
  \overline{g}_{\rho\nu}\overline{\nabla}_{\mu} \xi^{\rho}+\overline{g}_{\rho\mu}\overline{\nabla}_{\nu} \xi^{\rho} + {\cal O}(h)\,,
\end{split}
\end{equation}
which is often called \emph{full quantum symmetry}.
Indices can now be raised and lowered by the background metric. Formula~\eqref{eq:hmunu-transf-linear} is given
to the order that is necessary for our computation,
which does not happen to contain $\lambda$. However, higher loop computations need
additional orders in $h_{\mu\nu}$, so we show two additional orders and their $\lambda$-dependence in
App.~\ref{sect:nonlinear-fluctuations}.

As gauge-fixing we choose a familiar de Donder form
\begin{equation}\label{eq:dedonder}
\begin{split}
 S_{\rm gf} &= \frac{1}{2\alpha} \int {\rm d}^dx \sqrt{\overline{g}}\,\overline{g}^{\mu\nu} F_\mu F_\nu\,,
 \\
 F_\mu & = \overline{\nabla}^\rho h_{\rho\mu} - \frac{\beta}{2} \overline{g}^{\rho\theta}\overline{\nabla}_\mu h_{\rho\theta}\,,
\end{split}
\end{equation}
which includes two gauge fixing parameters $\alpha$ and $\beta$. Notice that this gauge fixing only changes the
fluctuation's two point function of the theory, while higher vertices
are unaffected by the gauge fixing.
Alternatively one could use the metric $g_{\mu\nu}$ instead of the background in the construction of \eqref{eq:dedonder}
at the price of modifying further vertices and making the ghosts action more complicated.
Speaking of which, the ghost action for ghost $c^\mu$ and antighost $\overline{c}^\mu$ fields is computed straightforwardly
\begin{equation}\label{eq:ghost}
\begin{split}
 S_{\rm gh} &= \int {\rm d}^dx \sqrt{\overline{g}} \, \overline{c}^\mu \left.\delta_\xi F_\mu\right|_{\xi\to c} \,.
\end{split}
\end{equation}
A common choice for the gauge fixing parameters is $\alpha=\beta=1$, because it allows to write
the fluctuation's Hessian as an operator of Laplace-type and to express everything in terms of simple heat kernel formulas.
Computations with arbitrary $\alpha$ and $\beta$ become difficult very quickly.

However, we still want to test gauge independence of our results, so in this section we make the simple choice
\begin{equation}\label{eq:gf-choice}
\begin{split}
 \alpha=1 \,, \qquad \beta=1+\delta\beta \,,
\end{split}
\end{equation}
in which $\delta\beta$ in an infinitesimal parameter. The advantage of this choice is that we still have
a simple Hessian because we can expand in powers of $\delta\beta$ and compute (for example) the first order correction
in the gauge dependence of our computations. The disadvantage is that the first order only proves that we are in a saddle point
of the gauge dependence, and does not prove full gauge independence. We thus assume that the cancellation of linear contributions
in $\delta\beta$ is an evidence for gauge independence.
In practice, the linear contributions in $\delta\beta$ emerge as new vertices with two external legs in the fluctuations
and in the ghosts, coming respectively by inserting \eqref{eq:gf-choice} in \eqref{eq:dedonder} and \eqref{eq:ghost},
which dress the two point functions.

For the computation we adopt the formalism of the effective action, which includes the one loop radiative corrections
\begin{equation}\label{eq:gamma}
\begin{split}
 \Gamma &= S_E + \frac{1}{2} \Tr \log {\cal O}_h - \Tr \log {\cal O}_{\rm gh}\,,
\end{split}
\end{equation}
coming from the Hessian operators, ${\cal O}$, of metric fluctuations and ghosts. For our needs it is sufficient
to compute $\Gamma$, which is naively a separate function of $\overline{g}_{\mu\nu}$ and $\langle h_{\mu\nu}\rangle$,
at $\langle h_{\mu\nu} \rangle=0$. Since we use dimensional regularization and have no explicit cutoff, we assume
that the split symmetry introduced by \eqref{eq:linear-split} is not broken
and that the corresponding Ward identity can be used to reconstruct immediately the full $\Gamma$.
Divergences are computed covariantly with heat kernel methods \cite{Avramidi:2000bm,Jack:1983sk,Martini:2018ska},
so that the theory can be renormalized as a standard field theory in curved space \cite{Brown:1980qq}.
We follow the three steps procedure
outlines in Sect.~\ref{sect:c}, the result is expressed in terms of the background
\begin{equation}\label{eq:sct}
\begin{split}
 \Gamma_{\rm div} =& 
  \frac{\mu^{-\zeta}}{4\pi\zeta}\int {\rm d}^d x \sqrt{\overline{g}}\Bigl\{
  \frac{g_0}{g_1} \Bigl[
  -\frac{1}{2} d (d+1)
  + \frac{d (d^2-d-4) \lambda }{4 (d-2)}
  \\&
  -\delta\beta  \Bigl(d-\frac{d \lambda  }{2-d}\Bigr)
  \Bigr]
  + \overline{R} \Bigl[
  \frac{5 d^2-3d+24}{12}
    \\&
  + \frac{1}{4} (-d^2+d+4) \lambda +\delta\beta  (d+\lambda -2)
  \Bigr]
  \Bigr\}
\,,
\end{split}
\end{equation}
in which $\overline{R}=R[\overline{g}]$ and it is important to remember
that it was computed at $\langle h_{\mu\nu} \rangle=0$.
The regularization to $d=2-\zeta$ of momentum space (heat kernel) integrals introduces the scale $\mu$,
that should be regarded as a momentum scale in reference to the background $\overline{g}_{\mu\nu}$
and that balances the change in the dimensionality of ${\rm d}^2x\to {\rm d}^dx$ and which we use for the RG.
As one might expect, the divergence proportional to the volume is also proportional to $g_0$,
because $g_0$ is the only parameter in the theory with the correct dimensionality, but also to $(g_1)^{-1}$,
because the perturbative coupling of the theory is Newton's constant $G=(g_1)^{-1}$.
Less expected could be the gauge and parametric dependence
that is displayed through $\delta\beta$ and $\lambda$, respectively.

In principle, following Ref.~\cite{Jack:1983sk},
one could eliminate \eqref{eq:sct} by redefining the bare parameters $g_0$ and $g_1$ of \eqref{eq:einstein-hilbert},
resulting in the coefficients of the poles becoming RG beta and gamma functions of the two couplings.
However, both functions would be gauge and parametric dependent, implying that they cannot be
associated to a physical observable of the theory.
The reason why this happens is that, out of the two monomials appearing in \eqref{eq:einstein-hilbert},
only one is independent on-shell, and therefore the information of only one physical beta function can be extracted.
To put \eqref{eq:sct} on-shell we choose the following parametrization
\begin{equation}\label{eq:sct-comp}
\begin{split}
 \Gamma_{\rm div} &
  = \frac{\mu^{-\zeta}}{\zeta} \int {\rm d}^d x \sqrt{\overline{g}} \Bigl\{
  A \overline{R} + J_{\mu\nu}  \Bigl( \overline{G}^{\mu\nu} + \frac{g_0}{2g_1}\overline{g}^{\mu\nu}\Bigr)\Bigr\}
\,,
\end{split}
\end{equation}
for some scalar constant $A$, a symmetric tensor $J_{\mu\nu}$ coupled to the equations of motion
of \eqref{eq:einstein-hilbert}, and the background's Einstein's tensor
$\overline{G}_{\mu\nu}=\overline{R}_{\mu\nu}-\nicefrac{1}{2} \overline{R} \,\overline{g}_{\mu\nu}$.
On-shell the equations of motion of the bare action vanish at this order, and therefore any dependence on $J_{\mu\nu}$
decouples \cite{Falls:2017cze}, while $A$ remains and should be subtracted
with a counterterm of Newton's constant \cite{Kawai:1989yh,Hamber:2007fk}.
We stress that with this choice of parametrization the volume operator has, by construction,
a trivial scaling.\footnote{%
A completely equivalent choice is to use the volume operator as first monomial of \eqref{eq:sct-comp}
so that it is the cosmological constant that provides a countertem, while Newton's constant would scale trivially.
We choose Newton's constant because it is more naturally the perturbative coupling of the theory.
}
Comparing \eqref{eq:sct} and \eqref{eq:sct-comp}, we find
\begin{equation}\label{eq:sct-coeff}
\begin{split}
 A&=\frac{36+3d-d^2}{48 \pi }\,,
\\
 J_{\mu\nu} &= \frac{\overline{g}_{\mu\nu}}{4\pi} \Bigl\{
 \frac{d^2-d-4 }{2 (d-2)}\lambda-\delta\beta  \left(2+\frac{2 \lambda
   }{d-2}\right)-d-1
\Bigr\} \,.
\end{split}
\end{equation}
The important result is that the coefficient $A$ is gauge and $\lambda$ independent,
so we can subtract the associated divergence and expect a physically sound interpretation,
while the tensor $J_{\mu\nu}$ carries the unwanted dependencies.\footnote{%
The procedure that we describe is discussed in other works. See, for example,  Refs.~\cite{Hamber:2007fk,Falls:2017cze,Kallosh:1978wt}.
However, we seem to have a different result as for the gauge-dependence of Ref.~\cite{Hamber:2007fk}.
}

Subtracting the first term of \eqref{eq:sct-comp} through a countertem for $g_1$ leads to the beta
function for Newton's constant $\beta_G = -A G^2$. According to the third step discussed in Sect.~\ref{sect:c}
we now also continue the theory above two dimensions replacing $G \to G \mu^{-\epsilon}$
which results in a scaling term for $\beta_G$.
Finally, we have
\begin{equation}\label{eq:betaG-JJ}
\begin{split}
 \beta_G&=\epsilon G - \frac{36+3d-d^2}{48 \pi } G^2\,,
\end{split}
\end{equation}
while, according to our choice of parametrization in \eqref{eq:sct-comp}, $g_0$ is not renormalized,
so the volume operator scales trivially \cite{Falls:2017cze}.\footnote{%
An equivalent choice would be to parametrize the first term of \eqref{eq:sct-comp} without the curvature scalar,
which would result in a nontrivial gamma function for $g_0$ and a trivial one for $g_1$.
This can be understood as a different choice of units. We recommend the discussion of Ref.~\cite{Percacci:2004yy} on this point.
}
Eq.~\eqref{eq:betaG-JJ} concides with the beta function given in \cite{Falls:2015qga},
which was obtained directly in $d>2$ with a cutoff, therefore we confirm that it is a universal result.
Some additional comments are in order.
An interesting aspect of \eqref{eq:betaG-JJ} is that
it gives perturbative information in $G$, which results in a perturbative information in $\epsilon=d-2$
at the RG fixed point. However, upon identification of the two parameters, 
the dimensionality $d$ appears also parametrically through the coefficient $A$
(recall that we regard this $d$ dependence as the $N$ dependence of an $SU(N)$ gauge theory).

The limit $d\to 2$ of \eqref{eq:betaG-JJ} agrees with Refs.~\cite{Jack:1983sk,Souma:1999at}
\begin{equation}\label{eq:betaG-JJ-2d}
\begin{split}
 \beta_G&=-\frac{19}{24\pi} G^2\,.
\end{split}
\end{equation}
A scaling relation suggests that
$\beta_G= \nicefrac{c}{24\pi} G^2$ with $c$ a number known as the central-charge in the context of CFT
\cite{Codello:2014wfa}, suggesting $c_{\Diff}=-19$.
Even in absence of a conformal interpretation, the value $-19$ is well-known
in the perturbative renormalization of $2d$ gravity \cite{Jack:1990ey} and, from our point of view,
it is a signature of the Einstein's realization of ${\Diff}$. One interesting point to make is that
if we instead decided to take the gauge-dependent counterterms \eqref{eq:sct} at $\lambda=\delta\beta=0$
and renormalize \eqref{eq:sct} off-shell, we would still get $c_{\Diff}=-19$ in the same limit,
but also a beta function for $g_0$, as it is commonly done in the literature of asymptotic safety.
This limit, which is gauge dependent because a change in $\beta$
would result in a different $\beta_G$, by \emph{accident} gives the on-shell physical value \cite{Jack:1983sk}.

Rather surprisingly, for $d=2$, $\lambda=0$ and with the gauge choice $\delta\beta=0$,
the beta function for $G$ obtained from the off-shell counterterms \eqref{eq:sct}
is the same as the correct on-shell one given later in \eqref{eq:betaG-JJ-2d}.
We argue that this unforeseen accident has lead many to believe
that the coefficient of the gauge- and $\lambda$-independent beta function \eqref{eq:betaG-JJ-2d}
was actually caused by the so-called linear parametrization $g_{\mu\nu}=\overline{g}_{\mu\nu}+h_{\mu\nu}$
($\lambda=0$ in \eqref{eq:linear-split}),
while it is actually independent on the parametrization \cite{Falls:2017cze}.\label{footnote:accident}

Another interesting aspect is to evaluate when $A>0$ as a function of $d$, which is the condition
for which $\beta_G$ has a nontrivial ${\cal O}(\epsilon)$ fixed point.
We find that the condition for the existence of a real fixed point returns a conformal window
$-4.685 < d < 7.685$ \cite{Falls:2015qga}.
This leads us to some interesting speculations: on the one hand, $d=4$ is well within the upper bound
of this fixed point; on the other hand, there is an \emph{effective upper critical dimension}
for this continuation, so the theory is not asymptotically safe up to $d\to \infty$, as suggested in
Ref.~\cite{Litim:2003vp}, but rather is safe up until $d_c=\nicefrac{1}{2} \left(3+3 \sqrt{17}\right)\approx 7.7$.
This is also one of the conclusions of Ref.~\cite{Gies:2015tca}, in which it is predicted $d_c \approx 5.7$.
Of course, the size of the conformal window is expected to change
by including further orders of perturbation theory as $d_c=d_c(\epsilon)$, which must also take into account
the nonlinear interplay between $d$ and $\epsilon$.
Nevertheless, it is an important benchmark for our argument to find agreement with
a sophisticate parametric analysis such as the one in Ref.~\cite{Gies:2015tca}, but also
a useful validation for  Ref.~\cite{Gies:2015tca} to find a perturbative argument confirming its analysis.

Having mentioned the possibility of going to two-loops order, there are some important aspects to clarify
in this regard. For the next-to-leading computation of $\Gamma$
it is not sufficient to renormalize all two-loop diagrams
at $\langle h_{\mu\nu} \rangle=0$ (see Ref.~\cite{Parker:2009uva} for a concise explanation).
This happens because ${\Diff}$ is realized nonlinearly from the point of view of $h_{\mu\nu}$, but also
because the split \eqref{eq:linear-split}, which is written in terms of bare quantities,
differs from its renormalized version.
In other words, the \emph{true quantum metric} $\langle g_{\mu\nu} \rangle$ of Einstein gravity
is different from the expectation value of the bare relation \eqref{eq:linear-split} and requires additional renormalization conditions.
To achieve a full two-loops computations, however, it is not necessary
to renormalize all possible vertices with external $h_{\mu\nu}$ legs, either.
The most economic way to go about is to renormalize a one-point function in $h_{\mu\nu}$,
which could be associated to a nonlinear source as done for the $2d$ nonlinear sigma model in Ref.~\cite{Howe:1986vm}.
This step is necessary to ensure that $\nicefrac{1}{\zeta^2}$ poles correctly cancel between two loop diagrams
and one loop diagrams with counterterms insertions \cite{Osborn:1987au}.
The cancellation of higher order poles is generally regarded as an important
nontrivial check in dimensional regularization, which unfortunately does not have a clear equivalent
in functional RG approaches to asymptotic safety.

A final comment concerns the conclusions of Ref.~\cite{Jack:1990ey}, in which it is argued that
it is impossible to renormalize the theory beyond the leading order, because of the kinematic pole.
The assumption made in Ref.~\cite{Jack:1990ey} is that one should find counterterms for both the $\nicefrac{1}{\zeta}$
poles of momentum diagrams and for the kinematic $\nicefrac{1}{d-2}$ poles, but this conspires against
the delicate cancellation of the $\nicefrac{1}{\zeta^2}$ poles. A similar argument could be made
using the one loop renormalization of curvature square composite operators, in which one might naively expect that
kinematic poles survive (because the relevant diagrams have more propagators than vertices)
and make the RG of some observables divergent in the limit $d\to 2$.
The problem with Ref.~\cite{Jack:1990ey} is that the result is gauge dependent, and correctly reproduces
$c_{\Diff}=-19$ only accidentally.
The procedure that we have used to go on-shell clearly lifts away unwanted dependencies, such as gauge's, but also
kinematic pole's as evident from $J_{\mu\nu}$ of \eqref{eq:sct-coeff}.
Preliminary results by some of us \cite{to-appear}
show that this problem is lifted, at least for composite higher derivative operators.
It would be desirable, however, to check in detail the two loops renormalizability of the two dimensional
Einstein's action using the three steps prescription of Sect.~\ref{sect:c}.

\section{Unimodular-dilaton action}\label{sect:ak}

Now we concentrate on the unimodular-dilaton action \eqref{eq:unimodular-dilaton}.
The dilaton field $\varphi$ and the unimodular metric $\tilde{g}_{\mu\nu}$ must be varied independently.
Following the discussion of Sect.~\ref{sect:diff}, we admit a deviation from $\Weyl$ symmetry,
so, effectively, the theory could just be thought as the one of a scalar field coupled to a unimodular metric.
The hope is that at the critical point $\Weyl$ invariance is restored and combines with unimodularity
to produce a ${\Diffstar}$ invariant action.
Keeping this in mind, we again use the background field method
expanding $\varphi$ and $\tilde{g}_{\mu\nu}$
\begin{equation}\label{eq:exp-split}
\begin{split}
 \tilde{g}_{\mu\nu} &= \left({\rm e}^{\hat{h}}\right)_\mu{}^\rho \hat{g}_{\rho\nu}
 = \hat{g}_{\mu\nu} +\hat{h}_{\mu\nu} +\frac{1}{2} \hat{h}_{\mu\rho} \hat{h}^{\rho}{}_\nu +{\cal O}(h^3)\,,
 \\
 \varphi &= \varphi_0 + \chi
\,.
\end{split}
\end{equation}
The unimodular metric $\tilde{g}_{\mu\nu}$ is expanded using a fluctuation $\hat{h}_{\mu\nu}$
which is traceless over the unimodular background $\hat{g}_{\mu\nu}$,
that is, $\hat{h}_\mu{}^\mu=\hat{g}^{\mu\nu}h_{\mu\nu}=0$.
With this choice the volume form is preserved both by infinitesimal and finite transformations,
$\sqrt{\tilde{g}}={\rm e}^{\tr \hat{h}}\sqrt{\overline{g}}=\sqrt{\hat{g}}$.

We stress the comparison between \eqref{eq:exp-split} and \eqref{eq:linear-split}: in the unimodular-dilaton
case we are forced to adopt an exponential form
(otherwise we would be breaking the $\Diffstar$ symmetry of the background)
and the volume element of the metrics $\tilde{g}_{\mu\nu}$ and $\hat{g}_{\mu\nu}$
are forced to be the same.
The first line of \eqref{eq:exp-split} can also be interpreted as the finite transformation relating
two unimodular metrics in the same equivalence class. When integrating over the field $\hat{h}_{\mu\nu}$, it is important to
keep in mind that it is traceless, for example when contracting the identity in its field space.

The dilaton $\varphi$ instead is expanded about a background $\varphi_0$, which is
not necessarily constant, though we ultimately will specialize to a constant dilaton to obtain the beta functions.
In fact, the size of $\varphi_0$ changes the normalization of the operator associated
to $g_1$ in the unimodular theory, $\tilde{R}$,
and this becomes important when evaluating the renormalization group scale.
Adopting a redefinition of the dilaton field, it is easy to see that $\varphi$ and $\tilde{g}_{\mu\nu}$
are the same degrees of freedom that are considered in Ref.~\cite{Aida:1996zn},
in which the configuration $\varphi_0=1$ is chosen to evaluate the counterterms.
Perhaps a bit controversial is the fact that we claim
that in Ref.~\cite{Aida:1996zn} the authors renormalized two dimensional unimodular-dilaton gravity,
rather than Einstein's realization, as should be evident from the conformal separation on the metric
and the discussion of Sect.~\ref{sect:diff}.
The motivation of Ref.~\cite{Aida:1996zn} was to match the results coming from $2d$ gravity \cite{Knizhnik:1988ak},
Liouville gravity and string theory \cite{Distler:1988jt,Kawai:1989yh}.
We return on this point with some more details later in Sect.~\ref{sect:d2}.

The discussion on background and true gauge symmetries of Sect.~\ref{sect:jj} applies equally well
to the background split \eqref{eq:exp-split} of this section, so we do not repeat it in its entirety.
In short, the symmetry to gauge fix is the one in which
the fluctuations, $\hat{h}_{\mu\nu}$ and $\chi$, transform as \eqref{eq:diff-unimodular-dilaton-action} at fixed backgrounds
$\hat{g}_{\mu\nu}$ and $\varphi_0$.
We gauge fix this action of ${\Diffstar}$ as in \eqref{eq:diff-unimodular-dilaton-action} with the gauge fixing
\begin{equation}\label{eq:dedonder-udg}
\begin{split}
 S_{\rm gf} &= \frac{1}{2\alpha} \int {\rm d}^dx \sqrt{\hat{g}} \, \hat{g}^{\mu\nu} F_\mu F_\nu
 \,,
 \\
 F_\mu &= \varphi_0 \, \hat{g}^{\nu\rho}\hat{\nabla}_\nu \hat{h}_{\rho\mu} - 2 \beta \hat{\nabla}_\mu \chi \,,
\end{split}
\end{equation}
and quantities with a hat come from the unimodular background metric.
This gauge fixing is chosen to reproduce \eqref{eq:dedonder} in the limit in which the metric fluctuations
of Sect.~\ref{sect:jj} and of this section are identified to the linear order.
What we mean is that if we combine $\chi$
and $\hat{h}_{\mu\nu}$ of this section in a symmetric tensor $h_{\mu\nu}$, then \eqref{eq:dedonder} and \eqref{eq:dedonder-udg}
are the same, and they also lead to the same leading ghost action (up to ${\cal O}(h)$ interactions) because \eqref{eq:linear-split}
and \eqref{eq:exp-split} are the same at the linear level (up to ${\cal O}(h^2)$ interactions).
This choice is made to ensure the same cancellations in the same limit $\alpha=\beta=1$. For the computations we adopt
$\alpha=1$ and $\beta=1+\delta\beta$ with $\delta\beta$ small as in the previous section to test gauge dependence.
Notice that the gauge fixings \eqref{eq:dedonder} and \eqref{eq:dedonder-udg} are not the same beyond the first order.

The computation of the effective action is similar to \eqref{eq:gamma} but in order to apply heat kernel techniques we factor out a local functional determinant proportional to the dilatonic background, ${\rm Det} [\varphi(x)\delta(x-y)]$, via a reparametrization of the quantum fields. Moreover the computation 
has two traces for the $(\hat{h}_{\mu\nu},\chi)$
and ghosts sectors, but at the leading order the first trace decouples further,
so it requires three distinct contributions
\begin{equation}\label{eq:gamma-udg}
\begin{split}
 \Gamma &= S_U + \frac{1}{2} \Tr \log {\cal O}_{\hat{h}} +\frac{1}{2} \Tr \log {\cal O}_\chi - \Tr \log {\cal O}_{\rm gh}\,.
\end{split}
\end{equation}%
The divergent part of $\Gamma$ coming from the fluctuations of \eqref{eq:unimodular-dilaton}
distinguishes of two parts: the leading radiative correction coming from the operators of the first line, and a multiplicative renormalization
of the cosmological constant $V$ and the topological charge operators, which we implement at the linear order by treating it as
a composite (local) operator.

The computation requires a ``Wick'' rotation of the dilaton's fluctuations to imaginary space to get rid
of the conformal mode instability, $\chi \to {\rm i} \chi$ for $d>2$, which is consistent with Ref.~\cite{Hawking:1978jn}.
We stress, however, that there is no instability for $d<2$, which is where loop integrals are regulated.\footnote{%
We also stress that other options for dealing with the dilaton Hilbert space are being seriously considered
in recent years by Morris and collaborators \cite{Morris:2018mhd,Mitchell:2020fjy,Morris:2020blt,Kellett:2020mle}.
}
We temporarily postpone the renormalization of the composite operators (the second line of \eqref{eq:unimodular-dilaton}).
The result for the divergent part of \eqref{eq:gamma-udg} is
\begin{equation}\label{eq:sct-udg}
\begin{split}
 \Gamma_{\rm div}
 =&
 \frac{\hat{\mu}^{-\zeta}}{\zeta} \int {\rm d}^d x \sqrt{\hat{g}} \Big\{
 \frac{-d^3+3d^2+42d+12}{48 d \pi}\hat{R}\\
 &+\frac{(d-2)(d+10)}{24 d \pi}\delta\beta\hat{R}\\
 &-\frac{3d^3-6d^2-12d+16}{8 d \pi}\frac{\hat{\nabla}^2\varphi_0}{\varphi_0}\\
 &-\frac{(d-2)(3d^2-4)(2d^2+d-2)}{16d^3\pi}\delta\beta\frac{\hat{\nabla}^2\varphi_0}{\varphi_0}
 \Big\}\,.
\end{split}
\end{equation}
The renormalization group scale $\hat{\mu}$
is a momentum scale that refers to the unimodular background $\hat{g}_{\mu\nu}$,
which is not Einsteinian, so we adopted a different notation to distinguish it from $\mu$ of Sect.~\ref{sect:jj}.

Similarly to the previous section, we want to ensure that the counter terms are evaluated on-shell.
The variational principle can be applied to \eqref{eq:unimodular-dilaton} by varying both $\varphi$
and $\tilde{g}_{\mu\nu}$, so we make the ansatz
\begin{equation}\label{eq:sct-udg-comp}
\begin{split}
 \Gamma_{\rm div} &
  = \frac{\hat{\mu}^{-\zeta}}{\zeta} \int {\rm d}^d x \sqrt{\hat{g}} \Bigl\{
  B  \hat{R} + J \hat{e}
\Bigr\}
\,.
\end{split}
\end{equation}
We introduced only the tensor $\hat {e}$ representing the equations of motion of $\varphi$ evaluated at $\varphi_0$,
because at this order the equations of motion of the unimodular metric do not contribute
given that they are traceless by construction. We find the coefficients
\begin{equation}\label{eq:sct-udg-coeff}
\begin{split}
 B&=-\frac{11d^4-44d^3-78d^2+180d-72}{96\pi d(d-1)}\\
 &\!\!\!-\frac{(d-2)(18d^5-35d^4-132d^3+152d^2+48d-48)}{192\pi d^3(d-1)}\delta\beta\,,\\
 J&=-\frac{3d^3-6d^2-12d+16}{8\pi d}\\
 &-\frac{(d-2)(3d^2-4)(2d^2+d-2)}{16\pi d^3}\delta\beta\,.
\end{split}
\end{equation}

We are almost ready to define the renormalization group flow. This can be done by taking $\varphi_0$
constant, following Ref.~\cite{Aida:1996zn}. However, the size of the dilaton would change
the beta function of Newton's constant multiplicatively, which explains the choice $\varphi_0=1$
made in Ref.~\cite{Aida:1996zn}. There is nothing wrong in having
a renormalization group flow that depends on the size of $\varphi_0$,
however we believe that a more elegant solution comes from first noticing that the magnitude
of the scale $\hat{\mu}$ of this section refers to the unimodular background metric $\hat{g}_{\mu\nu}$.
In other words, we suggest to think at $\hat{\mu}^2$
as a momentum square $\hat{\mu}^2 \approx p^\alpha p^\beta \hat{g}_{\alpha\beta}$.
A better comparison with the results of Sect.~\ref{sect:jj} would then suggest the use of a traditional Einsteinian
metric, and we can construct one by combining the background dilaton and the background unimodular metric
into a new Einsteinian metric
\begin{equation}\label{eq:new-bg}
\begin{split}
 \overline{g}_{\mu\nu} = \varphi_0^{\nicefrac{4}{d-2}}\hat{g}_{\mu\nu}\,,
\end{split}
\end{equation}
which we use to define a new scale $\mu^2 \approx p^\alpha p^\beta \overline{g}_{\alpha\beta}
\approx \varphi_0^{\nicefrac{4}{d-2}} \hat{\mu}^2$.
Combining everything together and using $d=2-\zeta$, we find the relation
\begin{equation}\label{eq:new-bg-scale}
\begin{split}
 \mu^{-\zeta} \approx \varphi_0^{-2} \hat{\mu}^{-\zeta} \,.
\end{split}
\end{equation}
It is straightforward to see that, with respect to the new scale $\mu$,
the counterterms are in form the same as the bare action \eqref{eq:unimodular-dilaton}.
We thus choose $\mu$
as our scale for the renormalization group and, thanks to the fact that it comes
from an Einsteinian metric, we also believe that it makes more sense to compare
its effects with those observed in Sect.~\ref{sect:jj}.

The computation of the RG equations follows the previous section:
we continue the theory to $d=2+\epsilon$ dimensions with $G\to G\mu^{-\epsilon}$
to get the beta function
$
\beta_G = \epsilon G - B G^2
$
with $B$ given in \eqref{eq:sct-udg-coeff}. The problem with the coefficient $B$
is that it is gauge-dependent, which can be seen easily from the second contribution
containing $\delta\beta$. This is the first real obstacle that our approach encounters;
we notice, however, that the gauge-dependent contribution is precisely zero for $d=2$,
so we know that at least in the two dimensional case the result is undoubtedly physical.
We believe that the gauge dependence appears because the anomaly is well-defined
only in $d=2$, for example $\langle T\rangle \propto  R$ can only be integrated in $d=2$,
giving the nonlocal Polyakov action \cite{Polyakov:1987zb}. We make a point in this direction
in appendix~\ref{sect:topological}.

In the limit $d\to 2$ our result agrees with Refs.~\cite{Kawai:1993mb,Aida:1996zn}
\begin{equation}\label{eq:betaG-ak2d}
\begin{split}
 \beta_G&=-\frac{25}{24\pi}G^2\,,
\end{split}
\end{equation}
which hints at the interpretation $c_{\Diffstar}=-25$ although, as we show in Sect.~\ref{subsect:aktojj}, the value of $c_{\Diffstar}$ is scheme dependent.
This is the result that one would expect from a quantization of $2d$ gravity ``a-la-string"
and it can be used to predict the critical dimension of the string worldsheet.
We expand on this point further in Sect.~\ref{sect:d2}.

In this scheme, we also have counterterms for the topological charge and the cosmological constant operators, which we give in
appendix~\ref{sect:topological}. 
Nevertheless, the RG of the volume operator becomes trivial in $d=2$ as we see from the beta functional for the dimensionless cosmological constant,
defined as the functional $\lambda(\varphi)\equiv\mu^{-d}V(\mu^{d_\varphi}\varphi)$:
\begin{equation}\label{eq:betaV-ak}
\begin{split}
 \beta_\lambda&= -d \lambda(\varphi) +d_\varphi\varphi \lambda'(\varphi) + \frac{d-2}{16\pi d} G \lambda''(\varphi)\,,
\end{split}
\end{equation}
where we have adopted the general dimension $d$
for later use, and denoted $d_\varphi=(d-2)/2$ the canonical dimension of $\varphi$, which is interpreted as a scalar field.
The reason why we have a separate beta function for the cosmological constant is that
in constructing the action \eqref{eq:unimodular-dilaton} we actually went outside the symmetry ${\Diffstar}$
by giving to the dilaton a functional dependence in the cosmological constant operator.
To discuss the beta function of $\lambda(\varphi)$ in \eqref{eq:betaV-ak} notice first that
the fixed functional ordinary differential equation, $\beta_\lambda=0$, is linear, so it is not
going to ``fix'' a critical value for $g_0$.
In the weak coupling limit (Gau{\ss}ian theory), the solution is simply $ \lambda(\varphi) \sim \varphi^{\nicefrac{2d}{d-2}}$,
thus reproduces the expected limit, which can be seen in \eqref{eq:dilaton} upon identification
of the overall constant with $g_0$, $ \lambda(\varphi) = g_0 \varphi^{\nicefrac{2d}{d-2}}$.
In this sense, the volume operator $\int\sqrt{g}$ scales trivially, in analogy with Sect.~\ref{sect:jj},
although the actual relation between the two schemes is detailed in Sect.~\ref{subsect:aktojj}.
General perturbative solutions to $\beta_\lambda=0$ have been investigated in Ref.~\cite{Aida:1994zc},
so we do not repeat them here.

We conclude this section by trying to infer some nonperturbative information
on the spectrum of solutions using an extension of \eqref{eq:betaV-ak}.
Notice first that a similar equation has already appeared in the renormalization of
$2d$ scalar potentials in Ref.~\cite{ZinnJustin:1989mi} although with a different overall sign
of the radiative correction, and is known to be related to nonperturbative information
on the Sine-Gordon potential (in particular the Coleman phase, see the third appendix of Ref.~\cite{Codello:2017hhh}).
As first step, we change the subtraction in \eqref{eq:betaV-ak} by rewriting the bare coupling $G_B$
as $\mu^\epsilon G_B=24\pi \epsilon G/(24\pi \epsilon-25 G)$, where we used
the values $d=2$ of \eqref{eq:sct-udg} for simplicity. We then take $\epsilon=d-2$,
because we want to be able to see the ${\cal O}(\epsilon)$ fixed point,
and get
\begin{equation}\label{eq:betaV-ak-2}
\begin{split}
 \beta_\lambda&= -d \lambda(\varphi) +d_\varphi \varphi \lambda'(\varphi) + \rho(d,G) \lambda''(\varphi)\,,
\end{split}
\end{equation}
where we defined
\begin{equation}\label{eq:rho}
\begin{split}
 \rho(d,G) &= \frac{(d-2)^2}{16\pi d}\frac{G}{ (d-2)-\frac{25}{24\pi}G}\,.
\end{split}
\end{equation}
The function $\rho(d,G)$ has two different signs according to the value of $G$ being either left or right
of the $G^*\sim {\cal O}(\epsilon)$ fixed point, which can be interpreted as the two phases of quantum gravity \cite{Kawai:1992np}.
For $G<G^*$ one finds $\rho>0$ and $G\to 0$ by RG evolution, where the theory presumably runs into
the Gau{\ss}ian phase and interpolates with the effective theory of gravity \cite{Donoghue:1994dn};
instead for $G>G^*$ one finds $\rho<0$ and $G\to \infty$ by RG evolution, so the theory is driven to
the strongly interacting phase (the analogy is with high- and low-$T$ phases in ferromagnets).

Irrespectively of the phase, we can study the spectrum of fluctuations by expanding
\begin{equation}\label{eq:lambda-exp}
\begin{split}
 \lambda(\varphi) &= \lambda^*(\varphi) + Y(\varphi) \left(\frac{\mu}{\mu_0}\right)^{-\theta}\,,
\end{split}
\end{equation}
where $\lambda^*(\varphi)$ is the solution of $\beta_\lambda=0$ with arbitrary conditions, $\mu_0$ is a reference scale,
and $\theta$ is the critical exponent of the deformation $Y(\varphi)$.
Using \eqref{eq:lambda-exp} in \eqref{eq:betaV-ak-2} we find a Schr\"odinger-like equation for $Y(\varphi)$
which can be solved as a Sturm-Liouville problem. Both phases have continuous and discrete spectrum,
similarly to the Halpern-Huang potentials \cite{Halpern:1994vw,Halpern:1995vf},
but the continuous part of the spectrum can be declared unphysical following
standard arguments (see, for example, Ref.~\cite{Bridle:2016nsu}).
We are left with a discrete spectrum for both sides, $\theta_n=d-n d_\varphi$ for the Gau{\ss}ian phase,
and $\theta_n=d-2n d_\varphi$ for the strongly-interacting one, given $n \in \mathbb{N}$.
It is easy to see that $n=0$ reproduces the expected scaling of the volume operator
and bounds the spectrum of scaling dimensions $\Delta_n=d-\theta_n$ from below;
there is thus a finite number of relevant directions on both sides,
in agreement with the general requirements of asymptotic safety.

\subsection{Relation between schemes}\label{subsect:aktojj}

As explained in Ref.~\cite{Falls:2017cze}, it is possible to recover the value of the central charge we found in the ${\Diff}$ via a suitable renormalization scheme, which we do in the gauge $\delta\beta=0$ for simplicity.
If we consider the equations of motion coming from the action \eqref{eq:dilaton}, we can choose which operator is renormalized and which operator ends up having a trivial (classical) scaling on-shell. Starting from the computation that was carried out in
Sect.~\ref{sect:jj}, our ansatz \eqref{eq:sct-comp} can be employed for the ${\Diffstar}$ case as well by solving the equations of motion for the coupling $g_0$.
In fact, summing \eqref{eq:sct-udg} for $\delta\beta=0$ to \eqref{eq:composite-operators} with $V''(\varphi)= \frac{2d(d+2)}{(d-2)^2}g_0\varphi^\frac{4}{d-2}$ and imposing
\begin{equation}
\begin{split}
g_0 =\frac{ \varphi^{-\frac{d+2}{d-2}}}{d G}\left[(d-2)\tilde{R}\varphi-4(d-1)\nabla^2\varphi\right]\,,
\end{split}
\end{equation}
then it is straightforward to get the on-shell divergence
\begin{equation}
\begin{split}
\Gamma_{\rm div} =&
	\frac{\hat{\mu}^{-\zeta}}{\zeta} \int {\rm d}^d x \sqrt{\hat{g}} \Bigl\{
	\frac{36+3d-d^2}{48\pi}\hat{R}\\
&-\frac{3d^4-12d^3-4d^2+36d-24}{8d(d-2)\pi}\frac{\hat{\nabla}^2\varphi}{\varphi}\Bigr\}\,,
\end{split}
\end{equation}
which reproduces the beta function \eqref{eq:betaG-JJ} for the Newton's constant through the coefficient of $\hat{R}$.
As a consequence we get $c=-19$ in $d=2$ as in the $\Diff$ realization of Sect.~\ref{sect:jj}.
Interestingly, this formula also leads to a vanishing anomalous dimension for the dilaton field in two dimensions.

Note that the converse, i.e.\ obtaining the value $c=-25$ from Einstein's realization, is not possible since $\Weyl$ is not a subgroup of $\Diff$ and, therefore, there is no way of solving the equations of motion with respect to a $\Weyl$ invariant operator.
In this case it is necessary to select some external $\Weyl$-invariant operator,
for example a matter operator with conformal symmetry, and arrange the counterterms so that it does not scale
with the RG flow~\cite{Falls:2017cze}.

\section{Conformal mode and $\bm{d\to 2}$}\label{sect:d2}

In order to appreciate the difference between the Einstein and unimodular-dilaton realizations,
it is important to clarify the role of the conformal mode and the $d\to 2$ limit.
The gauge fixed two point function in $h_{\mu\nu}$ of the Einstein action \eqref{eq:einstein-hilbert} in the gauge $\alpha=\beta=1$
of \eqref{eq:dedonder} is
\begin{equation}\label{eq:2pf-einstein}
\begin{split}
 \frac{\delta^2}{\delta h_{\mu\nu}\delta h_{\rho\theta}} \Bigl(S_E+S_{\rm gf}\Bigr)
 &=
 -\nabla_x^2 \delta_{xy} K^{\mu\nu\rho\theta} +E^{\mu\nu\rho\theta}
\,,
\end{split}
\end{equation}
with $E^{\mu\nu\rho\theta}$ an endomorphism that depends on the curvatures, and a matrix $K$ defined as
\begin{equation}\label{eq:Kdef}
\begin{split}
 K_{\mu\nu\rho\theta} &= \frac{1}{2}\Bigl(g_{\mu\rho}g_{\nu\theta}+g_{\mu\theta}g_{\nu\rho}-g_{\mu\nu}g_{\rho\theta}\Bigr)\,.
\end{split}
\end{equation}
The Green function of \eqref{eq:2pf-einstein} requires the inverse of $K$, which becomes
\begin{equation}\label{eq:invKdef}
\begin{split}
 {K^{-1}}_{\mu\nu\rho\theta} &=
 \frac{1}{2}\Bigl(g_{\mu\rho}g_{\nu\theta}+g_{\mu\theta}g_{\nu\rho}-\frac{1}{d-2}g_{\mu\nu}g_{\rho\theta}\Bigr)\,,
\end{split}
\end{equation}
in which the kinematic pole at $d=2$ is evident from the third term.
The reason why there is a pole is because the number of degrees of freedom changes when ``crossing'' $d=2$ by analytic continuation\footnote{A similar discontinuity is also present in the generalized vector field theories where the nondegenerate model and the Abelian gauge theory can be related by a single parameter in the action, but their renormalization has to be carried out independently. We refer to \cite{Ruf:2018vzq} for a deeper discussion of these models in curved spacetime.}.
For arbitrary $d$ one expects a trace and a traceless degrees of freedom. Instead, at precisely $d=2$ each metric is locally conformally related to the flat metric. Naively one would expect
the factor $(d-2)$ to multiply the traceless part, so that it does not contribute, instead it appears as a pole magnifying the
trace (conformal) part. This happens because the Einstein contribution in \eqref{eq:einstein-hilbert} becomes a topological invariant at $d=2$.

In the past, the kinematic pole has been regarded as a serious problem for the quantization of $2d$ gravity,
because it was assumed that it should be treated as the poles coming from dimensional regularization and
consistently subtracted with counterterms \cite{Jack:1990ey}. This seems reasonable, but, upon further scrutiny,
the problem with this perspective is twofold.
One the one hand the dimensional poles of regularization must be subtracted with one scheme in mind,
for example in the modified minimal subtraction scheme ($\overline{\rm MS}$) one subtracts finite factors of $\pi$ and $\gamma$,
while such arbitrariness is not allowed for the kinematic pole.
On the other hand, higher order dimensional poles are constrained to be consistent with structural equations, so any modification
emerging from adding further poles likely breaks the constraints.
Specifically, at $L$-loops $\nicefrac{1}{\zeta^L}$ poles arise, but they must combine with lower loop counterterms to cancel
in the logarithmic RG derivative.

It is easy to discuss the case $L=2$: first recall that the RG scale $\mu$ comes paired with each new pole as $\mu^{L\zeta}$.
The constraint is that higher poles come in the combination
$\nicefrac{\mu^{2\zeta}}{\zeta^2}-2\nicefrac{\mu^\zeta}{\zeta^2}$,
where the first monomial is a genuine two loop divergence (hence $\mu^{2\zeta}$),
while the second is the product of a counterterm (giving one inverse power of $\zeta$) and a one loop
divergence (hence $\mu^\zeta$) \cite{Osborn:1987au}.
The relative coefficient between the two terms must be $-2$ in order to cancel the logarithmic derivative,
but this is simply not the case if one requires the subtraction of kinematic poles too,
which is essentially the argument of Ref.~\cite{Jack:1990ey}.
We have shown in Sect.~\ref{sect:jj} how kinematic poles are cancelled if one considers
only on-shell gauge independent quantities for the renormalization, arguing that $d$ dependence
can be treated like the parametric $N$ dependence for $SU(N)$ gauge theories.
However, we did not explicitly check that the aforementioned counterterm structure is
realized beyond one loop.
Certainly, it would be desirable
to have an explicit two loop renormalization of \eqref{eq:einstein-hilbert} that uses our three steps procedure
to completely dismiss the claims of two loop nonrenormalizability of
Ref.~\cite{Jack:1990ey}.

A different perspective on how to address the same ``problem'' was developed through several papers many years ago,
culminating in the result of Ref.~\cite{Aida:1996zn} with the explicit motivation of realizing the original idea of
Kawai and Ninomiya \cite{Kawai:1989yh} that $2d$ gravity a-la-string
should arise as the limit $\epsilon\to 0$ of a theory of gravity in $d=2+\epsilon$ (see also \cite{Odintsov:1991qu, Odintsov:1991nf, Elizalde:1995km, Elizalde:1995gw} for related works).
To clarify what ``a-la-string'' means: in the modern version of the path integral of string theory,
one generally switches from the Nambu-Goto to the string Polyakov's action
introducing an auxiliary metric. For consistency, it is necessary to ensure that the conformal mode
of the string is not quantized, nor it is propagating because of an anomaly,
so one generally adopts a conformal gauge-fixing that allows to integrate the auxiliary metric
only over nonconformal degrees of freedom. This results in a ghost system, generally known as $bc$-system,
that contributes $c_{bc}=-26$ to the conformal anomaly.

One way to think at this contribution to the anomaly is that it should come from a full integration
of all metric degrees of freedom, if one somehow could also eliminate the trace mode effective action too.
In practice, the interpretation is to see the anomaly $c_{bc}=-25-1$, in which $-1$ is a contribution of
a \emph{subtracted} conformal mode effective action, and $-25$ is the expected central charge of $2d$ gravity.
For obvious reasons, this program can only work if conformal and nonconformal (the $bc$-system) are
separated in the construction of the path integral. By construction, a separation procedure circumvents the problems
of \eqref{eq:invKdef} because the two modes can be rescaled independently so to eliminate the kinematic pole.
This is basically what has been done in the works leading to Ref.~\cite{Aida:1996zn},
in which it has been shown how to obtain the $-25$ contribution to the anomaly.

The interpretation that we suggest in this paper is that Ref.~\cite{Aida:1996zn}
does not quantize $2d$ gravity a-la-Einstein \eqref{eq:einstein-hilbert},
but rather they do quantize unimodular-dilaton gravity \eqref{eq:unimodular-dilaton}, implying $c_{\Diffstar}=-25$.
A simple way to see this is to notice that the transformation \eqref{eq:diff-unimodular-dilaton-action}
are precisely the action of the transformation group shown in Ref.~\cite{Aida:1996zn},
but we show that they realize ${\Diffstar}$ instead of ${\Diff}$.
We stress that the two theories discussed in Sect.~\ref{sect:diff} are classically equivalent,
so there is absolutely nothing wrong with choosing the unimodular-dilaton realization over Einstein's.

A more sophisticate analogy comes from discussing the $d\to 2$ limit of \eqref{eq:einstein-hilbert},
which, on the basis of the above arguments, should interpolate both with a classical conformal anomaly action
and with the dilatonic action used in Ref.~\cite{Aida:1996zn}.
To see this, notice first that the would-be kinematic pole is ``hidden'' in the rescaling 
$g_{\mu\nu}=\varphi^{\nicefrac{4}{d-2}} \tilde{g}_{\mu\nu}$ which leads to the actions
\eqref{eq:dilaton} and \eqref{eq:unimodular-dilaton}.
Following partly Ref.~\cite{Nink:2015lmq}, we write $\varphi=1+\chi$, with the choice of the magnitude of the constant background
being arbitrary, and continue $d=2+\epsilon$ in \eqref{eq:unimodular-dilaton}
\begin{equation}
\begin{split}
 S_U[\psi,\tilde{g}] =& -g_1\int {\rm d}^d x \sqrt{\tilde{g}}\Bigl\{(1+\chi)^2 \tilde{R}
 + \frac{4}{\epsilon} \tilde{g}^{\mu\nu}\partial_\mu\chi\partial_\nu\chi\Bigr\}
 \\&
 + g_0 \int {\rm d}^d x \sqrt{\tilde{g}}\Bigl\{ (1+\chi)^{\nicefrac{4}{\epsilon}} \Bigr\} +{\cal O}(\epsilon)
\,,
\end{split}
\end{equation}
then we rescale $\chi$ to eliminate the poles $\chi = (\epsilon/8)^{\nicefrac{1}{2}}\psi$.
Taking the limit $\epsilon\to 0$, we notice that the rescaling becomes
$g_{\mu\nu}=\varphi^{\nicefrac{4}{d-2}} \tilde{g}_{\mu\nu} \to {\rm e}^{2\psi}\tilde{g}_{\mu\nu} $,
and we get
\begin{equation}\label{eq:dilaton-2d-action}
\begin{split}
 S_U[\psi,\tilde{g}] \to & -g_1\int {\rm d}^d x \sqrt{\tilde{g}} \Bigl\{
  (1+\psi) \tilde{R} + \frac{1}{2}\tilde{g}^{\mu\nu}\partial_\mu\psi\partial_\nu\psi\Bigr\} 
 \\&
 + g_0 \int {\rm d}^d x \sqrt{\tilde{g}}\Bigl\{  {\rm e}^{2\psi}\Bigr\} 
\,.
\end{split}
\end{equation}
The action of Ref.~\cite{Aida:1996zn} is obtained by applying the same steps, while
keeping the $\epsilon^{\nicefrac{1}{2}}$ and $\epsilon$ terms,
which results in a more complicate function of the dilaton, $L(\psi)=1+a\psi +b\psi^2$,
multiplying the curvature scalar (see also a pertinent discussion
in Ref.~\cite{Martini:2018ska}). The coefficients $a$ and $b$ in $L(\psi)$ are not independent,
so the renormalization is nonlinear and, beyond the leading order, requires a field redefinition of $\psi$
similarly to the $2d$ nonlinear sigma model \cite{Osborn:1987au}.

The action \eqref{eq:dilaton-2d-action} is clearly a Liouville action for $\psi$, but the interpretation of this paper is that
$\tilde{g}_{\mu\nu}$ is a \emph{dynamical} unimodular metric.
The Polyakov's anomaly action can be derived from \eqref{eq:dilaton-2d-action}
solving for $\psi$ to get a familiar nonlocal structure \cite{Polyakov:1987zb}, and one can clearly see that the Polyakov action induces a dynamics
for the conformal mode by explicitly multiplying the metric by a space dependent factor.

\section{Conclusions and further speculations}\label{sect:conclusion}

We have explored the renormalization of metric gravity with an Einstein-Hilbert-type action
in $d=2+\epsilon$ dimensions with the intent of framing the discussion on the four-dimensional
asymptotic safety conjecture from a different angle.
Quantum gravity in $d=2$ is asymptotically free, thus a simple dimensional argument shows that the beta function
of Newton's constant has a critical point in $d=2+\epsilon$, which could
represent a consistent ultraviolet completion of the theory and could circumvent the intrinsic limitations
of the effective field theory approach caused by perturbative nonrenormalizability in $d=4$.
Our aim is to develop a perturbative framework
that is reliable and gives benchmark limits for nonperturbative computations that are carried out directly in $d=4$.
There are several lessons that we draw from the perturbative computation in $d=2+\epsilon$, which we believe
should be carefully considered for the physical limit.

First and foremost, we have discussed how two different, but isomorphic, realizations of the diffeomorphisms group
lead to two theories of gravity that are classically equivalent, but quantum mechanically distinct.
The two realizations are different in the way in which the conformal transformations are treated:
in one case we have a more traditional theory described by a metric and an Einstein-Hilbert action,
while in the other case the metric is interpreted as the product of a conformal factor and another unimodular metric.
From the point of view of the renormalization, the two theories allow for different critical behavior
in the form of rather different beta function's coefficients, $c_{\Diff}=-19$ and $c_{\Diffstar}=-25$, respectively, the difference coming from the choice of operators one compares the running of Newton's constant to.
We have noticed that the first realization, $c_{\Diff}=-19$, has been incorrectly dismissed in the past,
even though it makes perfect sense, provided that only physical on-shell quantities are considered for the renormalization.
We have also noticed that the second realization, $c_{\Diffstar}=-25$, is precisely the one
that is generally associated to the string's worldsheet of string theory.
Importantly, the two realizations are different because of the intrinsic symmetry, and \emph{not}
because of a parametrization or gauge artifacts, as was sometimes believed.

We expect that a distinction between the two realizations of the diffeomorphisms group
should be made and also be very relevant for the nonperturbative approaches to quantum gravity
with metric degrees of freedom.
Another very important lesson comes from the necessity to go on-shell in order to have gauge and parametrization independence.
This is certainly not a surprise, but we have done it in such a way that, hopefully, can be generalized
to the nonperturbative approaches to quantum gravity and the asymptotic safety conjecture.
We find very reasonable that our results confirm an idea, also suggested by Nink and Reuter in Ref.~\cite{Nink:2015lmq},
that there are \emph{two distinct universality classes of quantum gravity}, which in this work appear
as the ${\Diff}$ and ${\Diffstar}$ realizations. The intriguing possibility, also pushed forward in Ref.~\cite{Nink:2015lmq},
is that this idea could be backed by explicit $2d$ conformal field theory results.
In fact, it was noticed before that $19$ is, among others, also a special value for the central charge \cite{Gervais:1990wv}.
It would be interesting to better understand this connection and find explicit realizations for the other special values.
One such realization could involve extrinsic information of the geometry \cite{Codello:2011yf}.

A naive question could be: why should we expect more than one universality class of quantum gravity?
As we have discussed, the two realizations are classically equivalent, but the path-integrals can be different.
The difference is therefore how the functional measure over the space of metrics, ${D g}$, is constructed.
Even though we do not think that framing this discussion on the different ``parametrizations'', exponential vs linear,
is the most appropriate way to address this problem, the conclusions of Ref.~\cite{Demmel:2015zfa}
still apply. The two expansions, \eqref{eq:linear-split} and \eqref{eq:exp-split},
end up integrating over two very different spaces of metrics, and there is absolutely no reason to believe
that these two integrations should produce the same results for the effective action and $\beta_G$.\footnote{An interesting perspective on the construction of the path integral measure in relation to the choice of frame in which the action is expressed can be found in \cite{Falls:2018olk}.}
The situation with Lorentzian metrics is probably even more complicate.

We believe that an important result of our work is the three steps procedure
to handle perturbative poles in quantum gravitational framework.
Our main idea is to distinguish the dimensionality $d$ coming from vertices and contractions of the metric,
which is an analog parameter to the $N$-dependence in $SU(N)$ gauge theory renormalization,
from the dimensionality $d=2-\zeta$ at which we regularize the theory, which leads to $\nicefrac{1}{\zeta}$ poles.
A crucial point is that we can have beta functions that depend on the first $d$ parametrically,
while having a well-behaved theory for $\zeta>0$.
Another crucial point is that, at the end, we continue the theory above $d=2$, meaning that we ``reverse''
the sign of $\zeta = -\epsilon <0$.
This last step is certainly the most delicate one, and it is the one that should be discussed
the most if one is willing to even attempt the continuation to $d=4$, which requires $\epsilon=2$.

Let us speculate a bit along this line.
The important question that we are left with is: can we continue the asymptotically free 
two-dimensional gravity models above their critical dimension and still have meaningful, possibly unitary, theories?
Our evidence suggests that, at least for infinitesimal $\epsilon$, this is always possible with minor parametric constraints on $d$,
but it would not be surprising if some serious obstruction emerges at finite values of $\epsilon$.
A similar situation has been discussed recently in a series of papers, culminating in Ref.~\cite{Giombi:2019upv},
in which the $O(N)$ model was continued above $d=4$. Ultimately, it was realized that for $4<d<6$
the $O(N)$ model is \emph{not} unitary because of instantons; critical quantities have exponentially small nonperturbative
imaginary contributions, leading to a complex conformal field theory interpretation in $d>4$.
The classification of gravitational instantons started a long time ago, see for example Ref.~\cite{Gibbons:1979xm}, and the outcome is
much more complicate than for scalar models.
However, there are now compelling reasons to include them in the path integral for metric gravity
and possibly in the discussion of the asymptotic safety conjecture \cite{Dvali:2010ue}.

Another speculation that we have suggested is the presence of an effective upper critical dimension,
implying that the continued theory to $d=2+\epsilon$, unitary or not, does not exist all the way to $d\to \infty$,
but rather it is nontrivial up to some $d_{cr} \gtrsim 4$.
This value seems to be consistently above $d=4$ for both realizations of the diffeomorphisms group considered in this work
(though the results for the $\Diffstar$ group are gauge dependent and consequently less reliable).
This is evidenced by the presence of two conformal windows for the coefficients of the beta functions continued in $d$ \cite{Falls:2015qga},
which, at least qualitatively, agree with previous findings by Gies et al.\ in Ref.~\cite{Gies:2015tca}.
A reliable procedure, perturbative or not, to compute such $d_{cr}$ would be very interesting because it could shine some
light on the underlying mechanism that actually produces a finite value for $d_{cr}$.
The logic that we imply here is that, forgetting the issue of unitarity, the theory could be continued
to finite values of $\epsilon$ from the perturbative point of view,
at least in principle, so there must be a nonperturbative interplay
with another entity that makes $d_{cr}$ finite.
Such interplay could be the collision of the $d=2+\epsilon$ fixed point with a multicritical
partner. An example of critical-multicritical collision in a simpler model
is the one between a critical and a multicritical fixed points,
which is believed to be responsible for the nontrivial form of the $q$-states Potts diagram
as a function of the parameters $(d,q)$ \cite{Codello:2020mnt}. We hope to come back to these topics in the future.

\smallskip

\paragraph*{Statement of contributions.}
The main results of the paper, presented in Sects.~\ref{sect:jj} and~\ref{sect:ak},
have been obtained mainly by AU and RM. The computation was checked both by hand
and with the Mathematica packages \emph{xAct} \cite{xact} and \emph{xTras} \cite{Nutma:2013zea}.
The discussions on symmetry and on the cosmological constant
operator presented in Sects.~\ref{sect:diff}~and~\ref{sect:ak} is based in part on the master thesis
of FDP presented at the University of Pisa in 2020.
OZ bears responsibility for any critique to the asymptotic safety programme
that appeared in the introduction and any complaint should be addressed to him exclusively.

\smallskip

\paragraph*{Acknowlegments.}
The research of AU was funded by the Deutsche Forschungsgemeinschaft
(DFG) under the Grant No.~Za 958/2-1.
OZ is grateful to several colleagues for many stimulating discussions over the years
on the topics covered in this paper,
but thanks especially A.~Nink and K.~Falls for comments and insights on earlier versiones of the draft.

\appendix

\begin{widetext}

\section{Cosmological constant and topological charge in UDG} \label{sect:topological}

We report here the results for the counterterms for two type of composite operators of the UDG action
\eqref{eq:unimodular-dilaton}.
They are a potential $V$ that depends on the dilaton and plays the role of cosmological constant,
and a coupling to the curvature  that generalizes the topological charge introduced in \eqref{eq:unimodular-dilaton}.
We thus introduce
\begin{equation}
\int {\rm d}^d x \sqrt{\tilde{g}}\Bigl\{ V(\varphi)+F(\varphi)\tilde{R} \Bigr\}\,.
\end{equation}
From the counterterms of the latter operators we can infer the beta function of the cosmological constant operator
given in \eqref{eq:betaV-ak}, and the renormalization of the topological charge discussed in Sect.~\ref{sect:diff}.

For the potential we find a divergence, which is used to derive \eqref{eq:betaV-ak}
\begin{equation}\label{eq:composite-operators}
\begin{split}
\Gamma_{\rm div}^{ V}&=-\frac{\hat{\mu}^{-\zeta}}{\zeta}\frac{d-2}{16\pi d g_1} \int {\rm d}^d x \sqrt{\hat{g}}\;V''(\varphi)\,.
\end{split}
\end{equation}
For the coupling to the curvature we find, generally, an off-shell divergence, that on-shell becomes
\begin{equation}\label{eq:composite-operators-counterterms-F}
\begin{split}
\Gamma_{\rm div}^{ F}&=-\frac{\hat{\mu}^{-\zeta}}{g_1\zeta}\int {\rm d}^d x \sqrt{\hat{g}}\;\Big\{\frac{(d-2)(d+2)(4d^4-24d^3-13d^2-36d+12)}{384 d^3 \pi}\frac{F(\varphi)}{\varphi^2}\\
		&+\frac{(d-2)(d+2)(d^2-d+2)}{128d^2\pi}\frac{F'(\varphi)}{\varphi}
		-\frac{(d-2)(d^2-2)}{16d^2\pi}F''(\varphi)\Bigr\}\hat{R}\,.
\end{split}
\end{equation}

The topological charge operator was introduced in the unimodular dilaton action \eqref{eq:unimodular-dilaton}
to compensate the quantum conformal anomaly with a classical term, so that the final result is Weyl-invariant.
The operator couples the scalar curvature to the dilaton field in a linear fashion in \eqref{eq:unimodular-dilaton}
\begin{equation}
\label{eq:top-charge-operator}
{\cal O}_{\rm top} = q \int {\rm d}^d x \sqrt{\tilde{g}} \,\varphi \tilde{R}\,,
\end{equation}
corresponding to the choice $F(\varphi)=q\varphi$ in \eqref{eq:composite-operators}.
A specific classical value of $q$ must be chosen to cancel the quantum contribution to the anomaly
as discussed in Sect.~\ref{sect:diff}.

In order to be able to set $q$ freely, we would like that $q$ itself, if seen as a coupling, would not run
with the renormalization group flow. This is however not the case:
if we renormalize \eqref{eq:top-charge-operator} as a composite operator we get from
\eqref{eq:composite-operators-counterterms-F}
a new on-shell contribution
 \begin{equation}
\begin{split}
\Gamma_{\rm div}^{\rm top}&=\frac{\hat{\mu}^{-\zeta}}{\zeta}\frac{q}{g_1}\int {\rm d}^d x \sqrt{\hat{g}} \Big\{
-\frac{(d-2)(d+2)(4d^4-21d^3-16d^2-30d+12)}{384d^3\pi}\frac{\hat{R}}{\varphi}
\Big\}\,.
\end{split}
\end{equation}
The obvious feature is that the subtraction of the on-shell divergence would imply a beta function for $q$
which is nonzero $\beta_q\neq 0$. This of course implies a nontrivial running for $\lambda$ and
the inability to eliminate the anomaly. However, for $d=2$ the counterterm is identically zero,
so $q$ can be set freely, in agreement with the fact that the gauge dependence disappears precisely in $d=2$.

Realistically, if for the conjecture of asymptotic safety requires the continuation to $d=4$
one could expect that it is the four dimensional conformal anomaly that actually must be canceled \cite{Riegert:1984kt},
eventually following the same steps discussed in Sect.~\ref{sect:diff}.
This is, arguably, difficult to achieve in the $\epsilon$-expansion,
but it might be possible with functional RG methods, since they work directly in $d=4$.

\end{widetext}

\section{On quadratic divergences in dimensional regularization} \label{sect:quadratic}

Sometimes, it is said that dimensional regularization with modified minimal subtraction, a.k.a.~$\overline{\rm MS}$,
does not depend on the scheme, which is not entirely true. Certainly, it
is unfair to other RG approaches such as functional RG \cite{Wetterich:1992yh}.
In Sect.~\ref{sect:introduction}, we have instead said that minimal subtraction is \emph{less} scheme dependent for two main reasons.
On the one hand, the leading and next-to-leading coefficients of the RG evolution equations
agree with all other approaches that do not feature an explicit cutoff scale, making them universal in this sense.
Other massless schemes would be ``simple'' minimal subtraction, but also, for example, traditional lattice perturbation theory.
On the other hand, it is often relatively easy to see explicitely that the RG equations of $\overline{\rm MS}$
are gauge-independent, implying that they could be related to
physical observables in the appropriate range of scales.
This happens because the analytic continuation of dimensional regularization does not break
the Ward-identities of gauge symmetry. In the background field approach used for this paper,
an explicit momentum cutoff would break the Ward identities introduced by splitting the metric as in \eqref{eq:linear-split},
and consequently the full quantum symmetry of $\Gamma$, given that the background symmetry is automatically preserved.

Obviously, the intrinsic price of dimensional regularization is that powerlaw divergences are lost in the analytic continuation,
unless one modifies the scheme to incorporate them as coming from lower dimensional poles
as was cleverly suggested by Jack and Jones \cite{Jack:1990pz,AlSarhi:1990np}.
This result could be an important starting point to revisit the asymptotic safety conjecture
with a more ``perturbative'' point of view.
In fact, most of the discussions on the asymptotic safety conjecture can be framed on whether the participants
to the debate do or do not \emph{believe} in quadratic divergences.

In this respect, perturbative methods employed at the critical dimension, such as those
of this paper, have the advantage of giving clear, controlled, and weakly scheme-dependent answers,
which might be more appropriate to settle this and other debates, if the results are compared
to the logarithmic divergences that are ``hidden'' in the functional RG approach.
A deeper and more correct interfacing between functional and dimensional methods might come
from an analysis such as the one of Ref.~\cite{Baldazzi:2020vxk}, in which a functional scheme is engineered that reproduces the results of dimensional regularization to an extent
(see also Ref.~\cite{Codello:2013bra} for an earlier discussion along similar lines).

\section{Nonlinear parametrization of quantum fluctuations} \label{sect:nonlinear-fluctuations}

Consider the rather general split transformation \eqref{eq:linear-split}, which breaks the full metric into
background and fluctuations. We recall it here for convenience
\begin{equation}\label{eq:linear-split-App}
\begin{split}
 g_{\mu\nu} &= \overline{g}_{\mu\nu} + h_{\mu\nu}
+\frac{\lambda}{2} h_{\mu\rho} \overline{g}^{\rho\theta} h_{\theta\nu} +{\cal O}(h^3)
\,.
\end{split}
\end{equation}
The considerations presented in this appendix apply equally well to the discussion of Sect.~\ref{sect:ak} with minor modifications.

As discussed in the main text, the split allows to manifestly preserve during every computation
a background version of the diffeomorphisms group, in which $h_{\mu\nu}$ behaves like a
standard symmetric tensor.
The action of the true diffeomorphisms group is defined as the transformation $\delta_\xi h_{\mu\nu}$,
which, if applied to \eqref{eq:linear-split-App}, reproduces $\delta_\xi g_{\mu\nu}= 2 \nabla_{(\mu}\xi_{\nu)}$
at fixed $\overline{g}_{\mu\nu}$.
We have that $\delta_\xi h_{\mu\nu}$ depends nonlinearly in $h_{\mu\nu}$, and the dependence can be
computed order by order.

We begin by acting with $\delta_\xi$ on \eqref{eq:linear-split-App}
\begin{equation}\label{eq:quantum-transformation-App}
\begin{split}
 g_{\nu\rho}\nabla_\mu \xi^{\rho}+g_{\mu\rho}\nabla_\nu \xi^{\rho} &=
 \delta_\xi h_{\mu\nu}
 +\lambda h_{(\mu}{}^\rho \delta_\xi h_{\nu)\rho}
 +{\cal O}(h^2)
\,,
\end{split}
\end{equation}
which must be solved in $\delta_\xi h_{\mu\nu}$.
We then introduce the following ansatz
\begin{equation}\label{eq:ansatz-deltah}
\begin{split}
\delta_\xi h_{\mu\nu} \equiv \delta_\xi^{(0)} h_{\mu\nu}
 + \delta_\xi^{(1)} h_{\mu\nu}
 + \delta_\xi^{(2)} h_{\mu\nu} + {\cal O}(h^3)\,, 
\end{split}
\end{equation}
in which $\delta_\xi^{(n)} h_{\mu\nu}$ is chosen to contain the tensorial contractions with $n$ powers of $h_{\mu\nu}$.
We plug the ansatz in \eqref{eq:quantum-transformation-App}
\begin{equation}\label{eq:quantum-transf-expansion-App}
\begin{split}
 &\xi^{\rho}\partial_{\rho}g_{\mu\nu}+g_{\nu\rho}\partial_\mu \xi^{\rho}+g_{\mu\rho}\partial_\nu \xi^{\rho} =
 \\
 & \qquad
 = \delta_\xi^{(0)} h_{\mu\nu} + \delta_\xi^{(1)} h_{\mu\nu}
  +\lambda h_{(\mu}{}^\rho \delta_\xi^{(0)} h_{\nu)\rho}
 \\
 & \qquad
 + \delta_\xi^{(2)} h_{\mu\nu}
  +\lambda h_{(\mu}{}^\rho \delta_\xi^{(1)} h_{\nu)\rho}
 +{\cal O}(h^3)
\,,
\end{split}
\end{equation}
in which we used the explicit form of the connection while keeping in mind that the metric on the left hand side must still be substituted with \eqref{eq:linear-split-App}.

The relation \eqref{eq:quantum-transf-expansion-App} can be solved order by order in the powers of $h_{\mu\nu}$
starting from $\delta_\xi^{(0)} h_{\mu\nu}$, which trivially has the form of the background diffeomorphism.
A dependence on the parameter $\lambda$ is introduced at first order by using the zeroth order solution in the
third term of the first line. Each order then contains polynomials with further powers of $\lambda$.
Explicitly, we find
\begin{equation}
\begin{split}
\delta_\xi^{(0)} h_{\mu\nu} &=
\overline{g}_{\rho\nu}\overline{\nabla}_{\mu}\xi^\rho +\overline{g}_{\rho\mu}\overline{\nabla}_{\nu}\xi^\rho
 \,,\\
 \delta_\xi^{(1)} h_{\mu\nu} &=
 \xi^\rho\overline{\nabla}_\rho h_{\mu\nu}
 -(\lambda-1)h_{\rho(\mu}\overline{\nabla}_{\nu)} \xi^\rho
 -\lambda \overline{\nabla}_\rho\xi_{(\mu} h_{\nu)}{}^\rho
 \,,\\
 \delta_\xi^{(2)} h_{\mu\nu} &=
 \frac{\lambda-1}{2}\xi^\rho\overline{\nabla}_\rho \left(h_{\mu\rho}h^\rho{}_\nu \right)
 +\frac{\lambda(\lambda+1)}{2} h_{\rho\sigma} h_{(\mu}{}^\rho \overline{\nabla}_{\nu)}\xi^\sigma
 \\
 &
 +\frac{\lambda(2\lambda-1)}{2} h_{\sigma(\mu} h_{\nu)}{}^\rho\overline{\nabla}_\rho \xi^\sigma 
 +\frac{\lambda^2}{2} h_\rho{}^\sigma\overline{\nabla}_\sigma\xi_{(\mu} h_{\nu)}{}^\rho
 \,,
\end{split}
\end{equation}
and so on, where indices on the right hand sides are raised and lowered using the background metric.
It is difficult to write down resummed formulas with arbitrary $\lambda$ for this expansion,
but it is possible to do it for the exponential form, that corresponds to $\lambda=1$, as shown in Ref.~\cite{Aida:1996zn}.



\begin{thebibliography}{99}
  
\bibitem{Goroff:1985th}
M.~H.~Goroff and A.~Sagnotti,
Nucl. Phys. B \textbf{266}, 709-736 (1986)

\bibitem{vandeVen:1991gw}
A.~E.~M.~van de Ven,
Nucl. Phys. B \textbf{378}, 309-366 (1992)

\bibitem{Donoghue:1994dn}
J.~F.~Donoghue,
Phys. Rev. D \textbf{50}, 3874-3888 (1994)
[arXiv:gr-qc/9405057 [gr-qc]].

\bibitem{Kawai:1989yh}
H.~Kawai and M.~Ninomiya,
Nucl. Phys. B \textbf{336}, 115-145 (1990)

\bibitem{Kawai:1992np}
H.~Kawai, Y.~Kitazawa and M.~Ninomiya,
Nucl. Phys. B \textbf{393}, 280-300 (1993)
[arXiv:hep-th/9206081 [hep-th]].

\bibitem{Codello:2014wfa}
A.~Codello and G.~D'Odorico,
Phys. Rev. D \textbf{92}, no.2, 024026 (2015)
[arXiv:1412.6837 [gr-qc]].

\bibitem{Polyakov:1987zb}
A.~M.~Polyakov,
Mod. Phys. Lett. A \textbf{2}, 893 (1987)

 \bibitem{Steinwachs:2020jkj}
 C.~F.~Steinwachs,
 [arXiv:2004.07842 [hep-th]].

\bibitem{Kawai:1993mb}
H.~Kawai, Y.~Kitazawa and M.~Ninomiya,
Nucl. Phys. B \textbf{404}, 684-716 (1993)
[arXiv:hep-th/9303123 [hep-th]].

\bibitem{Weinberg:1980gg}
S.~Weinberg,
In {\it General Relativity: An Einstein centenary survey}, 
ed.\ S.~W. Hawking and W. Israel, chapter 16, pp.790--831; 
Cambridge University Press.

\bibitem{Weinberg:1976xy}
S.~Weinberg,
In {\it Understanding the Fundamental Constituents of Matter. The Subnuclear Series}, vol 14,
ed.\ A.~Zichichi; Springer.

\bibitem{Bonanno:2020bil}
A.~Bonanno, A.~Eichhorn, H.~Gies, J.~M.~Pawlowski, R.~Percacci, M.~Reuter, F.~Saueressig and G.~P.~Vacca,
Front. in Phys. \textbf{8}, 269 (2020)
[arXiv:2004.06810 [gr-qc]].

\bibitem{Jack:1990ey} 
  I.~Jack and D.~R.~T.~Jones,
  Nucl.\ Phys.\ B {\bf 358}, 695 (1991).

\bibitem{Aida:1996zn} 
  T.~Aida and Y.~Kitazawa,
  Nucl.\ Phys.\ B {\bf 491}, 427 (1997)
  [hep-th/9609077].

\bibitem{Reuter:1996cp}
M.~Reuter,
Phys. Rev. D \textbf{57}, 971-985 (1998)
[arXiv:hep-th/9605030 [hep-th]].

\bibitem{Souma:1999at}
W.~Souma,
Prog. Theor. Phys. \textbf{102} (1999), 181-195
[arXiv:hep-th/9907027 [hep-th]].

\bibitem{Wetterich:1992yh}
C.~Wetterich,
Phys. Lett. B \textbf{301}, 90-94 (1993)
[arXiv:1710.05815 [hep-th]].

\bibitem{Souma:2000vs}
W.~Souma,
[arXiv:gr-qc/0006008 [gr-qc]].

\bibitem{Donoghue:2019clr}
J.~F.~Donoghue,
Front. in Phys. \textbf{8}, 56 (2020)
[arXiv:1911.02967 [hep-th]].

\bibitem{Falkenberg:1996bq}
  S.~Falkenberg and S.~D.~Odintsov,
  Int.\ J.\ Mod.\ Phys.\ A {\bf 13} (1998) 607
  [arXiv:hep-th/9612019[hep-th]].

\bibitem{Benedetti:2011ct}
D.~Benedetti,
New J. Phys. \textbf{14}, 015005 (2012)
[arXiv:1107.3110 [hep-th]].

\bibitem{Benedetti:2015zsw}
D.~Benedetti,
Gen. Rel. Grav. \textbf{48}, no.5, 68 (2016)
[arXiv:1511.06560 [hep-th]].

\bibitem{Nink:2015lmq}
A.~Nink and M.~Reuter,
JHEP \textbf{02}, 167 (2016)
[arXiv:1512.06805 [hep-th]].

\bibitem{Falls:2017cze}
K.~Falls,
Phys. Rev. D \textbf{96}, no.12, 126016 (2017)
[arXiv:1702.03577 [hep-th]].

\bibitem{Gielen:2018pvk} 
  S.~Gielen, R.~de Le{\'o}n Ard{\'o}n and R.~Percacci,
  Class.\ Quant.\ Grav.\  {\bf 35}, no. 19, 195009 (2018)
  [arXiv:1805.11626 [gr-qc]].

\bibitem{Demmel:2015zfa}
M.~Demmel and A.~Nink,
Phys. Rev. D \textbf{92}, no.10, 104013 (2015)
[arXiv:1506.03809 [gr-qc]].

\bibitem{Chernicoff:2018apt}
M.~Chernicoff, G.~Giribet, N.~Grandi, E.~Lavia and J.~Oliva,
Phys. Rev. D \textbf{98} (2018) no.10, 104023
[arXiv:1805.12160 [hep-th]].

\bibitem{Levy:2018bdc}
T.~Levy and Y.~Oz,
JHEP \textbf{06}, 119 (2018)
[arXiv:1804.02283 [hep-th]].

\bibitem{David:1988hj}
F.~David,
Mod. Phys. Lett. A \textbf{3} (1988), 1651
doi:10.1142/S0217732388001975

\bibitem{Kawai:1995ju}
H.~Kawai, Y.~Kitazawa and M.~Ninomiya,
Nucl. Phys. B \textbf{467}, 313-331 (1996)
[arXiv:hep-th/9511217 [hep-th]].

\bibitem{Aida:1994zc}
T.~Aida, Y.~Kitazawa, H.~Kawai and M.~Ninomiya,
Nucl. Phys. B \textbf{427}, 158-180 (1994)
[arXiv:hep-th/9404171 [hep-th]].

\bibitem{Martini:2018ska}
R.~Martini and O.~Zanusso,
Eur. Phys. J. C \textbf{79}, no.3, 203 (2019)
[arXiv:1810.06395 [hep-th]].

\bibitem{Distler:1988jt}
J.~Distler and H.~Kawai,
Nucl. Phys. B \textbf{321}, 509-527 (1989)

\bibitem{Hawking:1978jn}
S.~W.~Hawking,
NATO Sci. Ser. B \textbf{44}, 145 (1979)
PRINT-78-0745 (CAMBRIDGE).

\bibitem{Morris:2018mhd}
T.~R.~Morris,
JHEP \textbf{08}, 024 (2018)
[arXiv:1802.04281 [hep-th]].

\bibitem{Mitchell:2020fjy}
A.~Mitchell and T.~R.~Morris,
JHEP \textbf{06}, 138 (2020)
[arXiv:2004.06475 [hep-th]].

\bibitem{Morris:2020blt}
T.~R.~Morris,
[arXiv:2006.05185 [hep-th]].

\bibitem{Kellett:2020mle}
M.~Kellett, A.~Mitchell and T.~R.~Morris,
[arXiv:2006.16682 [hep-th]].

\bibitem{Gies:2015tca}
H.~Gies, B.~Knorr and S.~Lippoldt,
Phys. Rev. D \textbf{92}, no.8, 084020 (2015)
[arXiv:1507.08859 [hep-th]].

\bibitem{Avramidi:2000bm}
I.~G.~Avramidi,
Lect. Notes Phys. Monogr. \textbf{64}, 1-149 (2000)

\bibitem{Jack:1983sk}
I.~Jack and H.~Osborn,
Nucl. Phys. B \textbf{234}, 331-364 (1984)

\bibitem{Brown:1980qq} 
  L.~S.~Brown and J.~C.~Collins,
  Annals Phys.\  {\bf 130}, 215 (1980).
  
\bibitem{Hamber:2007fk}
H.~W.~Hamber,
[arXiv:0704.2895 [hep-th]].

\bibitem{Kallosh:1978wt}
R.~E.~Kallosh, O.~V.~Tarasov and I.~V.~Tyutin,
Nucl. Phys. B \textbf{137} (1978), 145-163

\bibitem{Percacci:2004yy}
R.~Percacci,
J. Phys. A \textbf{40}, 4895-4914 (2007)
[arXiv:hep-th/0409199 [hep-th]].

\bibitem{Falls:2015qga}
K.~Falls,
Phys. Rev. D \textbf{92}, no.12, 124057 (2015)
[arXiv:1501.05331 [hep-th]].

\bibitem{Litim:2003vp}
D.~F.~Litim,
Phys. Rev. Lett. \textbf{92}, 201301 (2004)
[arXiv:hep-th/0312114 [hep-th]].

\bibitem{Parker:2009uva}
L.~E.~Parker and D.~Toms,
``Quantum Field Theory in Curved Spacetime: Quantized Field and Gravity,''
Cambridge University Press (2009)

\bibitem{Howe:1986vm}
P.~S.~Howe, G.~Papadopoulos and K.~S.~Stelle,
Nucl. Phys. B \textbf{296}, 26-48 (1988)

\bibitem{Osborn:1987au}
H.~Osborn,
Nucl. Phys. B \textbf{294}, 595-620 (1987)

\bibitem{to-appear}
R.~Martini, A.~Ugolotti and O.~Zanusso,
to appear (2021)

\bibitem{Knizhnik:1988ak}
V.~G.~Knizhnik, A.~M.~Polyakov and A.~B.~Zamolodchikov,
Mod. Phys. Lett. A \textbf{3}, 819 (1988)

\bibitem{Peskin:1980ay}
M.~E.~Peskin,
Phys. Lett. B \textbf{94}, 161-165 (1980)

\bibitem{Gies:2003ic}
H.~Gies,
Phys. Rev. D \textbf{68}, 085015 (2003)
[arXiv:hep-th/0305208 [hep-th]].

\bibitem{Morris:2004mg}
T.~R.~Morris,
JHEP \textbf{01}, 002 (2005)
[arXiv:hep-ph/0410142 [hep-ph]].

\bibitem{ZinnJustin:1989mi}
J.~Zinn-Justin,
Int. Ser. Monogr. Phys. \textbf{77}, 1-914 (1989)

\bibitem{Codello:2017hhh}
A.~Codello, M.~Safari, G.~P.~Vacca and O.~Zanusso,
Eur. Phys. J. C \textbf{78}, no.1, 30 (2018)
[arXiv:1705.05558 [hep-th]].

\bibitem{Halpern:1994vw}
K.~Halpern and K.~Huang,
Phys. Rev. Lett. \textbf{74}, 3526-3529 (1995)
[arXiv:hep-th/9406199 [hep-th]].

\bibitem{Halpern:1995vf}
K.~Halpern and K.~Huang,
Phys. Rev. D \textbf{53}, 3252-3259 (1996)
[arXiv:hep-th/9510240 [hep-th]].

\bibitem{Bridle:2016nsu}
I.~Hamzaan Bridle and T.~R.~Morris,
Phys. Rev. D \textbf{94}, 065040 (2016)
[arXiv:1605.06075 [hep-th]].

\bibitem{Ruf:2018vzq}
M.~S.~Ruf and C.~F.~Steinwachs,
Phys. Rev. D \textbf{98} (2018) no.2, 025009
[arXiv:1806.00485 [hep-th]].

\bibitem{Odintsov:1991qu}
S.~D.~Odintsov and I.~L.~Shapiro,
Phys. Lett. B \textbf{263} (1991), 183-189
doi:10.1016/0370-2693(91)90583-C

\bibitem{Odintsov:1991nf}
S.~D.~Odintsov and I.~L.~Shapiro,
Int. J. Mod. Phys. D \textbf{1} (1992), 571-590
doi:10.1142/S0218271892000288

\bibitem{Elizalde:1995km}
  E.~Elizalde and S.~D.~Odintsov,
  Phys.\ Lett.\ B {\bf 347} (1995) 211
  [arXiv:hep-th/9501067[hep-th]].
  
  
\bibitem{Elizalde:1995gw}
  E.~Elizalde and S.~D.~Odintsov,
  Mod.\ Phys.\ Lett.\ A {\bf 10} (1995) 2001
  [arXiv:hep-th/9511031[hep-th]].

\bibitem{Gervais:1990wv}
J.~L.~Gervais,
Phys. Lett. B \textbf{243}, 85-92 (1990)

\bibitem{Codello:2011yf}
A.~Codello and O.~Zanusso,
Phys. Rev. D \textbf{83}, 125021 (2011)
[arXiv:1103.1089 [hep-th]].

\bibitem{Giombi:2019upv}
S.~Giombi, R.~Huang, I.~R.~Klebanov, S.~S.~Pufu and G.~Tarnopolsky,
Phys. Rev. D \textbf{101}, no.4, 045013 (2020)
[arXiv:1910.02462 [hep-th]].

\bibitem{Falls:2018olk}
K.~Falls and M.~Herrero-Valea,
Eur. Phys. J. C \textbf{79} (2019) no.7, 595
doi:10.1140/epjc/s10052-019-7070-3
[arXiv:1812.08187 [hep-th]].

\bibitem{Gibbons:1979xm}
G.~W.~Gibbons and S.~W.~Hawking,
Commun. Math. Phys. \textbf{66}, 291-310 (1979)

\bibitem{Dvali:2010ue}
G.~Dvali, S.~Folkerts and C.~Germani,
Phys. Rev. D \textbf{84} (2011), 024039
[arXiv:1006.0984 [hep-th]].

\bibitem{Codello:2020mnt}
A.~Codello, M.~Safari, G.~P.~Vacca and O.~Zanusso,
Phys. Rev. D \textbf{102}, no.12, 125024 (2020)
[arXiv:2010.09757 [cond-mat.stat-mech]].

\bibitem{xact}
J.~M.~Martín-García, \emph{xAct}, Efficient tensor computer algebra (2002),
\url{http://metric.iem.csic.es/Martin-Garcia/xAct/}

\bibitem{Nutma:2013zea}
T.~Nutma,
Comput. Phys. Commun. \textbf{185}, 1719-1738 (2014)
[arXiv:1308.3493 [cs.SC]].

\bibitem{Riegert:1984kt}
R.~J.~Riegert,
Phys. Lett. B \textbf{134}, 56-60 (1984)

\bibitem{Jack:1990pz}
I.~Jack and D.~R.~T.~Jones,
Nucl. Phys. B \textbf{342}, 127-148 (1990)

\bibitem{AlSarhi:1990np}
M.~S.~Al-Sarhi, D.~R.~T.~Jones and I.~Jack,
Nucl. Phys. B \textbf{345}, 431-444 (1990)

\bibitem{Baldazzi:2020vxk}
A.~Baldazzi, R.~Percacci and L.~Zambelli,
[arXiv:2009.03255 [hep-th]].

\bibitem{Codello:2013bra}
A.~Codello, M.~Demmel and O.~Zanusso,
Phys. Rev. D \textbf{90} (2014) no.2, 027701
[arXiv:1310.7625 [hep-th]].


  
\end{thebibliography}
\end{document}